\documentclass[11pt]{article}
\pdfoutput=1
\usepackage[mathscr]{euscript}

\DeclareSymbolFont{rsfs}{U}{rsfs}{m}{n}
\DeclareSymbolFontAlphabet{\mathscrsfs}{rsfs}
\usepackage{amsmath,amssymb,amsthm,mathrsfs,bbm,cite,MnSymbol,subfig}
\usepackage{graphicx}
\usepackage[pdftex,colorlinks=true, linkcolor={black}]{hyperref}
\usepackage[export]{adjustbox}
\usepackage[final,color]{showkeys} %%% should be loaded after hyperref        %%% add final to avoid printing the labels
\definecolor{refkey}{gray}{0.75}
\definecolor{labelkey}{RGB}{155,48,48}
\renewcommand*\showkeyslabelformat[1]{%
  \fbox{\parbox[t]{0.8\marginparwidth}{\raggedright\normalfont\scriptsize\url{#1}}}}

\allowdisplaybreaks
\allowbreak

\usepackage{IEEEtrantools}

\newcommand{\be}{\begin{equation}}
\newcommand{\ee}{\end{equation}}
\newcommand{\bea}{\begin{eqnarray}}
\newcommand{\eea}{\end{eqnarray}}
\numberwithin{equation}{section}

\newcommand{\ZZ}{{\rm ZZ}}
\newcommand{\FZZT}{{\rm FZZT}}

\newcommand{\xdownarrow}[1]{%
  {\left\downarrow\vbox to #1{}\right.\kern-\nulldelimiterspace}
}

\begin{document}
\begin{flushright}
\end{flushright}
~
\vskip5mm
\begin{center} 
%{\Large \bf ETH in SYK}
%{\Large \bf On Eigenstate Thermalization in the Schwarzian theory}
%\vskip10mm
{\Large \bf Extended Eigenstate Thermalization \\and the role of FZZT branes in the Schwarzian theory}
\vskip10mm

Pranjal Nayak\textsuperscript{a}, Julian Sonner\textsuperscript{b} \& Manuel Vielma\textsuperscript{b}\\
\vskip1em
\textsuperscript{a}Department of Physics \& Astronomy, University of Kentucky, 505 Rose St, Lexington, KY, USA\\
\textsuperscript{b}Department of Theoretical Physics, University of Geneva, 24 quai Ernest-Ansermet, 1211 Gen\`eve 4, Switzerland
\vskip5mm

\textsuperscript{a}\tt{pranjal.nayak@uky.edu}\\
\textsuperscript{b}\tt{\{julian.sonner, manuel.vielma\}@unige.ch}

\end{center}

\vskip10mm

\begin{abstract}
 In this paper we provide a universal description of the behavior of the basic operators of the Schwarzian theory in pure states. When the pure states are energy eigenstates, expectation values of non-extensive operators are thermal. On the other hand, in coherent pure states, these same operators can exhibit ergodic or non-ergodic behavior, which is characterized by elliptic, parabolic or hyperbolic monodromy of an auxiliary equation; or equivalently, which coadjoint Virasoro orbit the state lies on.  These results allow us to establish an extended version of the eigenstate thermalization hypothesis (ETH) in theories with a Schwarzian sector. We also elucidate the role of FZZT-type boundary conditions in the Schwarzian theory, shedding light on the physics of microstates associated with ZZ branes and FZZT branes in low dimensional holography.

\end{abstract}

\pagebreak
\pagestyle{plain}

\tableofcontents
%%%%%%%%%%%%%%%%%%%%%%%%%%%%%%%%%%%%%%%%%%%%%%
%%%%%%%%%%%%%%%%%%%%%%%%%%%%%%%%%%%%%%%%%%%%%%
%%%%%%%%%%%%%%%%%%%%%%%%%%%%%%%%%%%%%%%%%%%%%%
%%%%%%%%%%%%%  	NEW SECTION BEGINS 		%%%%%%%%%%%%%%
%%%%%%%%%%%%%%%%%%%%%%%%%%%%%%%%%%%%%%%%%%%%%%
%%%%%%%%%%%%%%%%%%%%%%%%%%%%%%%%%%%%%%%%%%%%%%
%%%%%%%%%%%%%%%%%%%%%%%%%%%%%%%%%%%%%%%%%%%%%%

%%%%%%%%%%%%%%%%%%%%%%%%%%%%%%%%%%%%%%%%%%%%%%
%%%%%%%%%%%%%%%%%%%%%%%%%%%%%%%%%%%%%%%%%%%%%%
%%%%%%%%%%%%%%%%%%%%%%%%%%%%%%%%%%%%%%%%%%%%%%
%%%%%%%%%%%%%  	NEW SECTION BEGINS 		%%%%%%%%%%%%%%
%%%%%%%%%%%%%%%%%%%%%%%%%%%%%%%%%%%%%%%%%%%%%%
%%%%%%%%%%%%%%%%%%%%%%%%%%%%%%%%%%%%%%%%%%%%%%
%%%%%%%%%%%%%%%%%%%%%%%%%%%%%%%%%%%%%%%%%%%%%%
\section{Introduction}
The Schwarzian theory, namely the theory defined by the path integral 
\be\label{eq.Schwarz}
 \int \mathscrsfs{D}\mu(f)e^{C\int \{ f(\tau), \tau\} d\tau}
\ee
where $\{f ; \tau \}$ is the Schwarzian derivative and $\mathscrsfs{D}\mu(f)$ is the appropriate measure on Diff$(S^1)$,
plays a key role in various different recent developments in low-dimensional holography. It appears as the low-energy description \cite{Maldacena:2016hyu} of the SYK model \cite{Sachdev:1992fk,Kitaev-talks:2015}, as well as related tensor models \cite{Gurau:2010ba,Witten:2016iux,Klebanov:2016xxf}. It can also be shown \cite{Jensen:2016pah,Maldacena:2016upp,Engelsoy:2016xyb, Mandal:2017thl} to arise in two-dimensional gravity theories, notably in the JT model \cite{Jackiw:1984je,Teitelboim:1983ux}. Furthermore it allows one to establish results on thermal properties of eigenstates and more general pure states in 2D CFT \cite{Turiaci:2016cvo,Anous:2016kss,Sonner:2017hxc,Anous:2017tza,Lam:2018pvp,Banerjee:2018tut,Vos:2018vwv,Anous:2019yku,Jensen:2019cmr}. In all these situations the appearance of the Schwarzian is intimately related to conformal (Virasoro) symmetry and its breaking \cite{Kitaev-talks:2015,Maldacena:2016hyu,Jensen:2016pah,Kitaev:2017awl}: an action of the type \eqref{eq.Schwarz} arises as the universal description of such effects in all instances mentioned.

In this paper we provide a universal description of the behavior of the basic operators of the Schwarzian theory in pure states. The philosophy behind our exploration of such effects is encapsulated in the eigenstate thermalization hypothesis (ETH) \cite{PhysRevA.43.2046,Srednicki}, reviewed for example in \cite{DAlessio:2016rwt}, and applied to the context of CFT in \cite{Lashkari:2016vgj,Lashkari:2017hwq}. In order to understand the thermalization of unitary closed quantum systems, this approach proposes to study the properties of eigenstates or typical pure states of the associated Hamiltonian and the degree to which operator expectation values in these states approximate those in thermal ensembles. The particular relevance of this approach to quantum thermalization in holography stems from the fact that the boundary dual of black hole formation and evaporation is one of thermalization in a closed quantum system, as demonstrated in detail for three-dimensional gravity in the series of papers \cite{Anous:2016kss,Anous:2017tza,Anous:2019yku}. 

Returning to the case at hand, namely theories of the type \eqref{eq.Schwarz} we find that their basic operators -- the bilocals  ${\cal O}(\tau_1, \tau_2)$  -- can have ergodic and non-ergodic behavior, i.e. they either approximate thermal ensembles well, or fail to thermalize, akin to the known behavior across a chaotic-integrable transition. Which behavior ensues depends on the parameters of the theory as well as the state, but is universally classified by elliptic, parabolic and hyperbolic monodromy, (or equivalently, which coadjoint Virasoro orbit the state lies on).  

This allows us to establish an extended version of the eigenstate thermalization hypothesis (ETH) \cite{Sonner:2017hxc,Anous:2019yku} which includes the usual notion of ETH for simple operators, and extends it to more complex operators, in particular including out-of-time-order correlation functions (OTOCs)\footnote{For an argument leading to a similar notion of extended ETH from the point of view of black-hole interiors, see \cite{deBoer:2019kyr}. }. Our argument proceeds along two closely related lines: firstly we give semi-classical arguments directly in the Schwarzian theory, based on the relationship between \eqref{eq.SchwarzCoherent} and the coadjoint orbit theory \cite{Alekseev:1988ce,Alekseev:1990mp}. Secondly we use the relationship between two-dimensional boundary Liouville theory and the Schwarzian theory to provide {\it exact} results, whose semi-classical expansion coincides with the previous analysis. A crucial novel ingredient in our story is the inclusion of effects of FZZT branes \cite{Fateev:2000ik,Zamolodchikov:2001ah} in the Liouville description, which descend to certain coherent states in the Schwarzian, as we shall explain. Effects of this type have recently been pointed out to arise in the study of the late-time behavior of a certain matrix model designed as a topological completion of JT gravity \cite{Saad:2019lba}. In this context FZZT type states are related to the late-time ramp behavior seen, for example, in the spectral form factor \cite{Cotler:2016fpe,delCampo:2017bzr,Altland:2017eao,Saad:2018bqo}. As we show here, these same effects are also seen to play an important role in understanding ETH in theories of the type \eqref{eq.Schwarz}.

This paper is structured as follows: the rest of this introduction consists of a summary of the main results. Section \ref{sec.FZZT} describes how to construct the Schwarzian path integral from two-dimensional Liouville theory. This kind of construction has already been exploited in \cite{Mertens:2017mtv,Lam:2018pvp}, but we extend it here to include more general boundary conditions that allow us to deal with a general class of pure states.  In section \ref{sec.Semiclassics} we then analyze the semiclassical limit of the correlation functions of interest and in particular provide a general proof that the heavy pure states of interest scramble as efficiently as the thermal ensemble, a result we refer to as {\it extended ETH}. Finally, section \ref{sec.ExactResults} makes full use of the construction of the Schwarzian theory in terms of boundary Liouville CFT to write down exact quantum expressions for the various pure-state expectation values considered in section \ref{sec.Semiclassics}. We also demonstrate that the exact expressions agree with semiclassics when expanded in that limit. We finally discuss our results and provide a perspective on open issues in the Conclusions section.

\subsection{Summary of results}\label{sec.SymmaryOfResults}
\begin{figure}[t]
\begin{center}
\includegraphics[width=0.8\columnwidth]{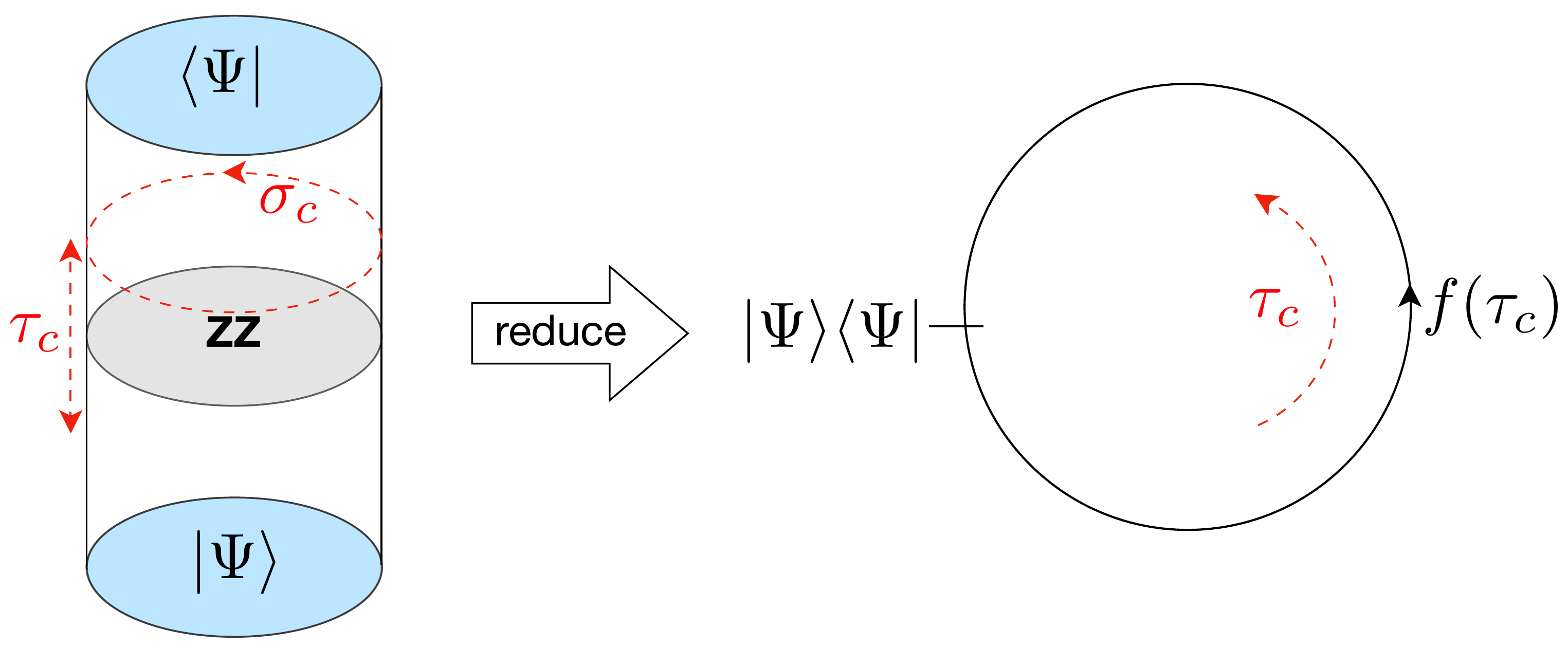}\\
\end{center}
\caption{We obtain results on Schwarzian correlation functions by taking an appropriate limit of 2D Liouville theory with boundary conditions, or equivalently (as shown on the left) with the corresponding boundary states (details in section \ref{sec.stateInterpretation}). The full cylinder is obtained when we implement the ZZ boundary condition via a doubling trick, allowing us to extend the field periodically. The resulting theory on the torus is then reduced to one dimension and gives rise to correlation functions in the Schwarzian theory with respect to the pure-state density operator $\rho_\Psi = | \Psi \rangle \langle \Psi|$.}
 \label{fig:SchwarzReduction}
\end{figure}
 
One of the main goals of this work is to establish results for correlation functions in pure states
\be\label{eq.HHnL}
{\rm Tr} \,\Bigl[ \rho_{\Psi} \,{\cal O}_{\ell_1}(\tau_1,\tau_1')\, \cdots\,  {\cal O}_{\ell_n}(\tau_n,\tau_n')    \Bigr]\,,\qquad \rho_\Psi = | \Psi\rangle \langle\Psi | 
\ee
where $|\Psi\rangle$ is either an eigenstate of the Schwarzian theory
\be\label{eq.Eigenstate}
H| E(k)\rangle = \frac{k^2}{2C}| E(k)\rangle  \,,
\ee
or a pure coherent state of the type
\be\label{eq.SchwarzCoherent}
| E_r \rangle \sim e^{ r{\cal V}} | E \rangle
\ee
where $|E\rangle$ is any eigenstate of the theory, including the vacuum $|0\rangle$. The operator ${\cal V}$ starts its life in two dimensions as a Liouville vertex operator $\sim e^{\varphi}$ that upon descent to one dimension corresponds to a bilocal operator insertion in the Schwarzian theory. We establish that for one and two-point functions of bilocal operators these correlation functions become thermal in the semiclassical limit ($C\rightarrow \infty$ ) in the sense of eigenstate thermalization and compute the associated ETH temperature $\beta_\Psi$. By definition, this temperature is also the temperature of a canonical ensemble with temperature chosen in order to reproduce microcanonical averages at energy $E$. We have written the result \eqref{eq.HHnL} for the case of an arbitrary number of bilocal insertions. Strictly speaking in this work we only establish the result for up to two bilocals, but our methods can in principle be extended to the most general case \eqref{eq.HHnL}.

 We will describe the precise construction in section \ref{sec.FZZT}, preferring to first summarize the main results in as non-technical a manner as possible.

\begin{figure}[t]
\begin{center}
\includegraphics[width=0.8\columnwidth]{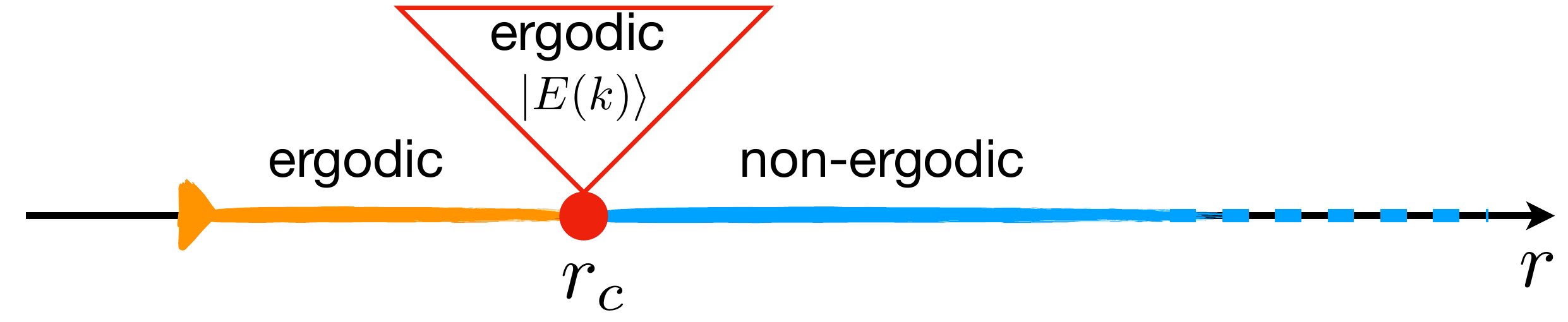}\\
\end{center}
\caption{Phase diagram of the chaotic properties of the Schwarzian theory. The theory behaves thermally for the parameter range $-\sqrt{2}<r<r_c$ with $r_c = \sqrt{2}$. The range $r<r_c$ corresponds to elliptic coherent state insertions, while the critical case $r=r_c$ can also be interpreted as the parabolic orbit corresponding to the eigenstates $|E(k)\rangle$ of the theory. In the full range $-\sqrt{2}<r\leq r_c$ we find that the model scrambles with the maximal Lyapunov exponent predicted by ETH temperature, $\lambda = 2\pi T_{\rm ETH}$. In the hyperbolic range $r>r_c$ the Schwarzian theory behaves non-ergodically, and in particular its OTOC is oscillatory. See also table \ref{fig.SummaryTable} for more details.}
 \label{fig:phaseDiag}
\end{figure}

We prove that both classes of pure states $|\Psi\rangle$ under consideration behave thermally to leading order in large $C$. We show both using semiclassical quantization of the appropriate coadjoint orbits of Virasoro as well as via semiclassical expansions of {\it exact} results in the quantum theory that bilocal one and two-point functions with respect to the eigenstate $|E(k)\rangle$ appear thermal at temperature $\pi T_E = \sqrt{E/2}$, while the coherent states $|E_{r}\rangle$ behave thermally at temperature 
\be\label{eq.CoherentEffectiveTemperature}
T_{\theta} = \frac{ \theta}{\pi}\sqrt{\frac{E}{2}} \,,
\ee
for $\theta\in\mathbb{R}$, corresponding to the parameter range $r<\sqrt{2}$ for the coherent state introduced in \eqref{eq.SchwarzCoherent} above.\footnote{$\theta$ and $r$ both parametrize the states and are related to each other through \eqref{eq.MonodromyTrace}.} These states behave non-ergodically for the parameter range $r>\sqrt{2}$, which formally corresponds to setting $\theta\in i \mathbb{R}$ in \eqref{eq.CoherentEffectiveTemperature}. In this regime there is no meaningful effective temperature to be associated to the states, but the combination of parameters \eqref{eq.CoherentEffectiveTemperature} still remains important, appearing for example as an oscillation frequency of  OTOCs. In fact, one recovers the result for the eigenstates by setting $\theta=1$, and we explain why in more detail below. Physically, the system shows a phase transition between ergodic and non-ergodic behavior at the critical value $r=\sqrt{2}$, very similar to the phenomenon established in \cite{Anous:2019yku} for holographic 2D CFT. The resulting phase diagram is shown in Figure \ref{fig:phaseDiag}. It is intriguing to see exactly the same mathematical structure at play here, namely the transition between the elliptic and hyperbolic monodromy, or equivalently the co-adjoint orbits of Virasoro.  The role of the critical theory is played by the parabolic case and it would be interesting to explore whether the theory of this orbit can serve as a universal description of ergodic to non-ergodic transitions.

 Note that we wrote the expectation value \eqref{eq.HHnL} with Euclidean time insertions, but our results also extend to Lorentzian insertions, and in particular to out-of-time order type correlation functions. In this case we are able to prove a conjecture made in \cite{Sonner:2017hxc}, namely that these pure states scramble with the maximally allowed Lyapunov exponent if one were allowed to naively extend the eigenstate thermalization hypothesis to these types of operators. In other words we find that
 \be
 \langle\Psi | {\cal O}_{\ell_1}(t,0) {\cal O}_{\ell_2}(t,0)  | \Psi\rangle_{\rm OTO} \sim 1 - \frac{\#}{C}e^{\frac{2\pi}{\beta_\Psi}t}
 \ee
up to the scrambling time.\footnote{An analogous statement is true in 2D CFT assuming identity block domination \cite{Anous:2019yku}. On general grounds, on the second sheet, one would expect contributions from other blocks to potentially spoil this behavior, unless we make further assumptions about the kind of CFT we consider, such as sparse spectrum and large gaps. In the Schwarzian theory the situation seems to be better as the identity block is all there is.} This behavior was conjectured in \cite{Sonner:2017hxc} on the basis of numerical evidence as well as in the works of \cite{deBoer:2018ibj,deBoer:2019kyr} with an eye on the necessary conditions for reconstructing the holographic bulk geometry of pure states. This kind of behavior has already been established analytically in 2D CFT assuming identity block domination in \cite{Anous:2019yku} and furthermore shown to be play a crucial role in holographic bulk reconstruction \cite{deBoer:2018ibj,deBoer:2019kyr}.

\subsection*{Extended eigenstate thermalization hypothesis}
A physical way to summarize these observations and analytical results, as well as previous evidence \cite{Sonner:2017hxc,Anous:2019yku}  is the statement that there exist theories satisfying an {\it extended} ETH. Let $\mathscrsfs{A}$ denote the set of operators which satisfy the ETH in its usual form \cite{PhysRevA.43.2046,Srednicki}. Then, by extended ETH, we mean that operators in eigenstates such as \eqref{eq.Eigenstate} and, as a consequence in pure states such as \eqref{eq.SchwarzCoherent}, approximate thermal ones in real time, up to exponentially small corrections in entropy, at least until the scrambling time $t_s$. Moreover we include in the set $\mathscrsfs{A}$  more complex operators, such as the four-point OTO type correlations considered in this paper as well as in \cite{Sonner:2017hxc,Anous:2019yku}. Furthermore, we suggest that certain theories saturate the bound \cite{Maldacena:2015waa} when the OTOC is evaluated in eigenstates, as was first conjectured in \cite{Sonner:2017hxc}, that is they satisfy {\it maximal extended ETH}.  One interesting class of such theories we have in mind are CFTs with a sparse spectrum and a large gap, $\Delta_{\rm gap}$, to higher spins \cite{Camanho:2014apa,Afkhami-Jeddi:2016ntf,Belin:2019mnx,Kologlu:2019bco}, in other words theories admitting a semiclassical description in terms of Einstein gravity corrected by higher derivative terms which are suppressed by powers of $\Delta_{\rm gap}$.  It would clearly be desirable to obtain a better understanding of how far one should be allowed to extend the admissible class operators, $\mathscrsfs{A}$, for example by adding successively higher OTOCs \cite{Haehl:2017qfl,Haehl:2017pak}. 

For demonstrating ETH in any system it is crucial that the off-diagonal expectation values of the operators under consideration, $\mathscrsfs A$, are exponentially suppressed with respect to the diagonal elements. Such off-diagonal terms were studied in \cite{Sonner:2017hxc, Nayak:2019khe}. In our present work we compute only the diagonal elements while the computation of off-diagonal terms will require generalizing some of the techniques developed in this paper.
\subsection*{Comments on ZZ and FZZT branes}
An interesting role in our story is played by ZZ and FZZT branes in the Liouville picture and the corresponding boundary conditions these imply for the Schwarzian path integral. A general lesson is that ZZ branes and FZZT branes enter somewhat asymmetrically, in so far as in the Schwarzian picture the ZZ branes give rise to integrals over the density of states, while FZZT branes allow us to consider non-trivial coherent external states with respect to which we compute expectation values.  On the other hand, we can recover the ZZ boundary conditions characterized by the continuous parameter $r$ rather simply, by choosing the single value $r=\sqrt{2}$, so that in practice all the associated pure states can be treated in a uniform fashion. As we shall see this provides a useful perspective on all these branes and their thermal properties by linking them to the different coadjoint orbits of Virasoro, again parametrized by $r$, with the critical case $r_c=\sqrt{2}$ (see, e.g. Fig. \ref{fig:phaseDiag}).

Given that these branes and their associated states feature so prominently in the understanding of thermalization in the Schwarzian theory (and by way of establishing this result, also in 2D Liouville theory), a very interesting question would be that of the bulk manifestation\footnote{The paper \cite{Mertens:2019tcm} discusses bulk manifestations of different Virasoro orbits, including elliptic, parabolic and hyperbolic, in terms of defect geometries in JT gravity.}. Since thermal ensembles in the boundary are a dual manifestation of bulk black holes, we are thus asking about the relevance of (F)ZZ(T) branes to the general structure of black-hole microstates. 

 In fact, in the context of the `old' $c\le 1$ matrix model many results in this direction are known \cite{Fateev:2000ik,Zamolodchikov:2001ah,Seiberg:2004at,Martinec:2003ka} and it would certainly be interesting to flesh out a similar story in the present case. For recent work in this direction see \cite{Saad:2019lba}.

\section{Pure states in Liouville and the Schwarzian}\label{sec.FZZT}
In this section we describe in detail how to construct expectation values with respect to a density operator $\rho_\Psi = |\Psi\rangle\langle \Psi |$ in the Schwarzian theory. On the way we also explain, in our picture, how to obtain the thermal ensemble $\rho_\beta$, more commonly encountered in the study of the Schwarzian theory to date. We will adopt the perspective advocated in \cite{Mertens:2017mtv} in order to do this, namely we will start with 2D Liouville theory with appropriate boundary conditions and then descend to the Schwarzian by taking an appropriate limit.

\subsection{Boundary Liouville theory and Schwarzian coherent states}\label{sec.BoundaryLiouville}
In this section we collect a number of results, most of which are well known, both to prepare the scene, as well as to establish our conventions for what is to come. We review how to descend from Liouville theory in 2D to the Schwarzian path integral \eqref{eq.Schwarz}, as well as the broad classification of classical solutions of Liouville according to their monodromy properties.  Let us start with the Hamiltonian path integral for boundary Liouville theory on the interval $I = [0,\beta/2]$,
\be\label{eq.boundaryLiouville}
\left\langle \quad\cdot\quad \right\rangle_{\rm bc} = \int \mathscrsfs{D}\varphi \mathscrsfs{D}\pi  \left(\,\, \cdot \,\, \right)e^{\int d\tau d \sigma \frac{\pi\dot\varphi}{8\pi b^2} - {\cal H}}
\ee
where `bc' generically denotes `boundary conditions' to be imposed on the fields at $\sigma = 0$ and $\sigma  = \beta/2$.\footnote{While it might seem counter intuitive to call the spatial direction $\beta$, the periodicity in this direction will eventually be related to Euclidean 1D time.} We shall be interested both in so-called ZZ branes as well as FZZT branes, which correspond to Dirichlet and Neumann conditions, respectively. The Hamiltonian density is given as
\be\label{eq.LiouvilleHamiltonian}
{\cal H} = \frac{1}{8\pi b^2} \left( \frac{\pi^2}{2} +  \frac{\varphi_\sigma^2}{2} +  e^{\varphi}   - 2 \varphi_{\sigma\sigma}  \right) + H_{\rm boundary}
\ee
with Liouville central charge $c = 1 + 6 \left(b + b^{-1}\right)^2$. We added the  boundary Hamiltonian $H_{\rm boundary}$ in order to implement the Dirichlet or Neumann conditions of interest and to be specified shortly. We will ultimately be interested in the limit $c\rightarrow \infty$, corresponding to taking $b\rightarrow 0$. The Schwarzian limit is essentially the classical limit of the 2D path integral \eqref{eq.boundaryLiouville} by taking $c\rightarrow \infty$ while at the same time letting $a\rightarrow 0$, such that $\frac{c\,a}{24 \pi} := C $ remains finite, where $a$ is the size of the Euclidean time coordinate in 2-dimensions. In this limit, the path integral reduces to the zero-mode in time of the fields $\varphi(\sigma,\tau)$ and $\pi(\sigma,\tau)$, i.e. time-independent configurations and may be brought into the form \eqref{eq.Schwarz}. A key difference to the treatment in \cite{Mertens:2017mtv,Lam:2018pvp,Mertens:2018fds} is that we are especially interested in FZZT type boundary conditions, so we describe in some detail how these are implemented and how they descend to the Schwarzian theory.

Let us thus return to our discussion of boundary Liouville theory. The boundary conditions we would like to impose \cite{Fateev:2000ik,Teschner:2000md,Zamolodchikov:2001ah,Dorn:2006ys} are most easily stated in terms of the vertex operator 
\be
{\cal V}_\ell(w,\bar w) = e^{\ell \varphi(w,\bar w)}\,,\qquad \left(w = \tau + i \sigma\,,\quad \bar w = \tau - i\sigma\right)
\ee
We also define the exponentiated Liouville field
\be
V(w,\bar w) = e^{-\frac{1}{2}\varphi}
\ee
for later convenience.The Schwarzian boundary states descend from considering the Liouville theory between branes, so that $\sigma\in I$. In order to characterize the branes, we need to specify Dirichlet or Neumann boundary conditions at both ends. The most general case of interest in this work\footnote{In section \ref{sec.ExactResults} we will also give results for Neumann boundary conditions at both ends. We do not currently understand the physical relevance of these states and leave further investigation for future work.} will be a Dirichlet condition at $\sigma=0$, as well as a family of Neumann boundary conditions at $\sigma = \beta/2$:
\be
V(\tau,\sigma )\Bigr|_{\sigma=0}=0\,,\qquad \partial_\sigma V(\tau,\sigma)\Bigr|_{\sigma=\frac\beta2} = \frac r2\,
\ee
where $r$ is a parameter whose ranges we will specify later. In terms of the field $\varphi$, the Neumann boundary condition takes the form
\be
\partial_\sigma \varphi + r e^{\frac{\varphi}{2}}\Bigr|_{\sigma=\frac\beta2} = 0
\ee
and is implemented by adding the boundary action
\be\label{eq.NeumannBdy}
S_{\rm N} =-\frac{r}{4\pi b^2} \int \!\!\!d\tau\, e^{\frac{1}{2}\varphi}\Biggr|_{\sigma = \frac\beta2}
\ee
which, inside the path integral, has precisely the interpretation of creating the coherent state \eqref{eq.SchwarzCoherent}. In preparation for our later analysis of the Schwarzian theory, it will in fact be useful to review the construction of classical solutions of boundary Liouville theory with the specified boundary conditions, and we closely follow the presentation in \cite{Dorn:2006ys}. The setup here is that of a conformal field theory with a boundary which was first studied in \cite{Cardy:1984bb} and subsequently in \cite{Zamolodchikov:2001ah, Fateev:2000ik, Teschner:2000md, Dorn:2006ys, Dorn:2008sw} for the Liouville field theory.  First, let us perform a conformal transformation that maps the strip, $\mathbb R\times I$, to upper half plane, $\mathbb H$,\footnote{Recall that the Liouville field $\varphi(w) \to \varphi(z) - \ln\left\vert\frac{\partial w}{\partial z}\right\vert^2$ also transforms under conformal transformations.}%
\begin{equation}\label{eq.StripToPlane}
	z=e^{\frac{2\pi}\beta w}, \quad \bar z = e^{\frac{2\pi}\beta \bar w}~.
\end{equation}
We will see below that this map naturally helps us to define a theory at finite temperature after the dimensional reduction (see the discussion following \eqref{eq.2DSchwarzian} below), although we will often be interested in taking the zero-temperature limit and instead work with individual pure states. Under such an exponential map, the boundaries $\sigma=0,\beta/2$ are mapped to the real line, $\textrm{Im}(z)=0$. The stress tensor on the plane is,
\be\label{eq.Fuchs}
T = \frac{\partial^2_z V}{V}+\frac1{4 z^2}\,,\qquad \bar T = \frac{\partial^2_{\bar z} V}{V} +\frac1{4 \bar z^2}~.
\ee
Conservation of the stress tensor implies holomorphicity and the Neumann boundary conditions naturally provide vanishing energy flux through the boundary, $\textrm{Im}(z)=0$,
\begin{equation}
	T(z) = \bar T(\bar z), \text{ when }z=\bar z
\end{equation}
For the case of Dirichlet boundary condition, this is imposed as an additional condition which imposes the regularity of the stress tensor on the boundary.\footnote{Non-vanishing energy flux for the Dirichlet problem would correspond to singularities on the boundary. See \cite{Dorn:2006ys} for more discussion. Moreover, this also facilitates the use of doubling trick.} The doubling trick on the plane then lets one define the stress tensor on the lower half plane, $T(z^*) = \bar T(\bar z)$, \cite{Cardy:1984bb}, thereby making the stress tensor periodic in the original $\sigma$-coordinate,
$$T(\sigma+\beta) = T(\sigma)~.$$
Classical solutions of the Liouville equation in the Fuchsian form \eqref{eq.Fuchs} are well studied and are organized by the monodromy of the solutions around the unit circle in the complex plane. To see this, we write the Liouville field as a linear combination of two functions  $\Psi^T = (\psi_1\,,\,\psi_2)$ in the form
\be
V(z,\bar z) = \Psi(\bar z)^T A \Psi(z)\,,\qquad A \in SL(2,\mathbb{R})
\ee
where the functions $\psi_{1,2}$ are defined as the two linearly independent solutions of Hill's  equation
\be\label{eq.HillsEquation}
\psi'' - T \psi = 0
\ee
with $T$ the Liouville stress tensor introduced above.  We then define the monodromy matrix $M$, via
\be
\Psi(e^{2i\pi}z) = M \Psi(z)\,.
\ee
One can show that $M\in SL(2,\mathbb{R})$, \cite{MW:Hills}, and that it is in fact defined only up to conjugation $M \cong S^{-1}MS$ for $S\in SL(2,\mathbb{R})$, so that conformally inequivalent solutions are labelled by conjugacy classes of $SL(2,\mathbb{R})$, coinciding with the classification of coadjoint orbits of Diff$(S^1)$ \cite{Witten:1987ty,Dorn:2006ys}. 
%\subsubsection*{Monodromy classification of solutions}
The main tool to classify the different classes of solutions is the monodromy matrix $M$, which must fall into one of three classes: hyperbolic, parabolic or elliptic,
characterized by the trace of the monodromy matrix $M$. We do this by writing the Neumann boundary condition in terms of $V(z,\bar z)$ as,
\be
	\frac{z\partial V(z,\bar z) -\bar z\bar \partial V(z,\bar z)}{\sqrt{z\bar z}}\Bigr|_{\bar z = z<0} = r/2~.
\ee%
With the aid of the Wronskian condition $\psi_1'\psi_2 - \psi_1\psi_2' = 1$, this implies that
\be\label{eq.MonodromyTrace}
{\rm Tr}M = \frac r2 = \frac{\cos(\pi\theta)}{\sqrt{2}}\,,
\ee
where the second equality introduces an alternative but standard parametrization of the boundary parameter $r$.
This equation will be our main tool to classify different types of semiclassical solutions of the Schwarzian model. Following the analysis in \cite{Dorn:2006ys}, we have three cases, namely 
\bea
-\sqrt2\ \ <\ \ r &<& \sqrt2\,\qquad \theta \in \mathbb{R}\,\qquad \quad \ \textrm{elliptic}\nonumber\\
r &=& \pm\sqrt2\,\quad\ \theta  = 0,\pm1\,\qquad \textrm{parabolic}\nonumber\\
r &>& \sqrt2\,\qquad \theta \in i\mathbb{R}\,\qquad \quad \textrm{hyperbolic}
\eea
Using the map \eqref{eq.StripToPlane}, one can write the classical solution for the Liouville theory found in \cite{Dorn:2006ys} in each of the equivalence classes in terms of a single holomorphic function, $F(z)$,
\begin{equation}
	V(z,\bar z) = \frac{1 }{2 \sqrt{2}  \theta } \sqrt{\frac{ F(z)F(\bar z) }{ F'(z) F'(\bar z) }} \left[\left( \frac{F(z)}{F(\bar z)} \right)^{\theta/2} - \left( \frac{F(\bar z)}{F( z)} \right)^{\theta/2} \right]~,
\end{equation}
where $F(e^{2i\pi}z) = e^{2i\pi}F(z)$.

A further crucial element of the construction of classical solutions described above is that the stress tensor of any solution of Hill's equation for the theory on $\mathbb{H}$ falling into these three classes can be brought to the constant form \cite{Dorn:2006ys}
\be\label{eq.constantOrbit}
T^\mathbb{H} = T_0^\mathbb{H} \,,\qquad \textrm{with} \qquad T_0^{\mathbb H} =\frac{\theta ^2}{4z^2} \,,
\ee
via a conformal transformation. This is the stress tensor on the plane. Alternatively, one can compute the stress tensor on the strip, $\mathbb R\times I$, where  it takes the form
\begin{equation}\label{eq.TconstantOrbit}
	T = T_0 = -\frac{\theta^2}4~.
\end{equation}
We refer to $T_0$ as the constant representative.\footnote{We have dropped the Casimir energy part that comes from the quantum corrections since it is not important for identifying the orbits.} It is interesting to note that this equation is very reminiscent of the trace of the monodromy matrix appearing in the computation of conformal blocks of 2D CFT at large central charge \cite{Zamolodchikov:1987ae,Hartman:2013mia,Anous:2016kss}. The difference seems to be in that while these references study the monodromy around operator insertions, here we have appropriate boundary conditions along the entire real line in the complex $z$-plane. However, following \cite{Fateev:2000ik} one can understand these boundary conditions as insertion of \emph{boundary} operators and the monodromy under study in the present work is then the monodromy around this boundary operator. 

\subsection{Descending to the Schwarzian}\label{sec.descent}
Inspired by the non-linear field transformation of Gervais-Neveu, \cite{GERVAIS198259, GERVAIS1982125}, we perform the following field redefinition in the path integral, \eqref{eq.boundaryLiouville},
\begin{equation}
\begin{aligned}\label{eq.GNdef}
	e^\varphi &= 8{\theta^2} \frac{ F'(z) F'(\bar z) }{ F(z)F(\bar z) } \left[\left( \frac{F(z)}{F(\bar z)} \right)^{\theta/2} - \left( \frac{F(\bar z)}{F( z)} \right)^{\theta/2} \right]^{-2}\\[5pt]
	\pi &= \frac{F''(z)}{F'(z)} + \frac{F''(\bar z)}{F'(\bar z)} - \frac{F'(z)}{F(z)} - \frac{F'(\bar z)}{F(\bar z)} -\theta \left[ \frac{F(z)^\theta+F(\bar z)^\theta}{F(z)^\theta-F(\bar z)^\theta} \right] \left[ \frac{F'(z)}{F(z)} - \frac{F'(\bar z)}{F(\bar z)} \right] ~.
\end{aligned}
\end{equation}
Under this redefinition, the bulk action becomes,
\begin{equation}\begin{aligned}
	S &= \frac1{8\pi b^2}\int\limits_{\mathbb H}\!\!d^2z \bigg[ \pi[F]\dot\varphi[F]+\left(2\{F(z);z\}+\left(1-\theta^2\right) \left(\frac {F'(z)}{F(z)}\right)^2 \right.\\
	&\hspace{5cm} + 2\{F(\bar z);\bar z\}+\left(1-\theta^2\right) \left(\frac {F'(\bar z)}{F(\bar z)}\right)^2\bigg) \bigg]
\end{aligned}\end{equation}
Using the doubling trick to identify $z^*=\bar z$, we write the action as an integral over the entire complex plane,
\begin{equation}\label{eq.2DSchwarzian}
	S[F] = \frac1{8\pi b^2}\int\limits_{\mathbb C}\!\!d^2z \Bigg[ \pi[F]\dot\varphi[F]+\Bigg(2 \{F(z);z\}+\left(1-\theta^2\right) \left(\frac {F'(z)}{F(z)}\right)^2 \Bigg) \Bigg]
\end{equation}
This path integral should be understood along with the insertion of the appropriate state along the negative real axis, implemented by the boundary term \eqref{eq.NeumannBdy}. Some comments are in order: firstly, recall that the $F(z)$ function is related to the transformations of the strip, $I$, by,
\be\label{eq.tan-transform}F(z) = e^{\frac{2\pi i}{\beta} f(\sigma)}\ee
on the $\tau=0$ slice.  The parameter $\beta$ entering into this transformation is freely tunable and corresponds to the physical (inverse) temperature of the Schwarzian theory in 1D. This is the well known $\tan$-transformation ($\mathbb{SL}(2,\mathbb R)$ equivalent thereof) from the study of 1D CFTs, \cite{Polchinski:2016xgd,Maldacena:2016hyu}. Secondly, note that $\theta=\pm1$ corresponds to the standard Schwarzian action. The reason for which will become clearer in Section \ref{sec.stateInterpretation}. Recall, that the semiclassical limit (of the Liouville theory), $b\to0$, of the path integral localizes on $\tau$-independent configurations and the first term ($\pi\dot\varphi$ term) in the above action drops out. In this limit, one reproduces the Schwarzian action,
\begin{equation}\label{eq.SchwarzianAfterDoubling}\begin{aligned}
	&\hspace{4cm}S[f] = \, -C\!\!\! \int\limits_{-\beta/2}^{\beta/2} \!\!d\sigma \Bigg[ \{f(\sigma);\sigma\}+ \frac{2\pi^2\theta^2}{\beta^2} f'(\sigma)^2 \Bigg]
\end{aligned}\end{equation}
where 
\be
C = \frac a{4\pi b^2}
\ee
 is the overall coefficient in front of the Schwarzian action \eqref{eq.Schwarz}.
Using the field redefinitions, \eqref{eq.GNdef}, insertion of any vertex operator, $e^{2\ell \varphi(z,\bar z)}$, corresponds to an insertion of the kind,
\begin{equation}\label{eq.OpBiLoc}
	\mathcal O_\ell(\sigma,-\sigma) = \left(\frac{8\pi^2\theta^2}{\beta^2}\frac{f'(\sigma)f'(-\sigma)}{\sin^2\left(\frac{\pi\theta}\beta \big(f(\sigma)-f(-\sigma)\big)\right)}\right)^{2\ell}
\end{equation}
in the classical limit.
Similarly, the boundary term \eqref{eq.NeumannBdy} can we written as,
\begin{equation}\label{eq.densityOp}
	S_N = -\frac{ a\,r \, \theta}{\sqrt2 \beta b^2} \ \frac{f'(\beta/2)}{\sin\left(\pi\,\theta\right)}
\end{equation}
where, we have used: $f(\beta/2)-f(-\beta/2) = \beta$; and, $f'(\beta/2) = f'(-\beta/2)$. From this point of view of obtaining Schwarzian theory as a dimensional reduction of the 2-dimensional Liouville theory, the Dirichlet boundary condition at $\sigma=0$ corresponds to an insertion of complete set of states in the Schwarzian theory.\footnote{Precisely because there is no insertion of any operator in the dimensionally reduced theory.} Finally, the generic 2-dimensional path integral, \eqref{eq.boundaryLiouville}, with operator insertions and Neumann boundary condition on one end and Dirichlet boundary condition on the other after dimensional reduction becomes,
\begin{equation}\label{eq.pathIntegralFinal}
	\langle\cdot\rangle = \int\!\!\frac{\mathscrsfs Df}{G} \Big(\cdot\Big) \exp\left[-\frac{ 2\sqrt2\pi C\,r \, \theta}{\beta} \ \frac{f'(\beta/2)}{\sin\left(\pi\,\theta\right)} -C \int\limits_{-\beta/2}^{\beta/2} \!\!d\sigma \Bigg(\{f(\sigma);\sigma\}+ \frac{2\pi^2\theta^2}{\beta^2} f'(\sigma)^2 \Bigg) \right]~.
\end{equation}
We can alternatively write the above integral in a more symmetric form:
\begin{equation}\label{eq.pathIntegralFinal_symm}
	\langle\cdot\rangle = \int\!\!\frac{\mathscrsfs Df}{G} \Big(\cdot\Big) \exp\left[-\frac{ 2\sqrt2\pi C\,r \, \theta}{\beta} \ \frac{\sqrt{f'(-\beta/2)f'(\beta/2)} }{\sin\left(\pi\,\theta\right)} -C \!\!\!\int\limits_{-\beta/2}^{\beta/2} \!\!\!d\sigma \Bigg(\{f(\sigma);\sigma\}+ \frac{2\pi^2\theta^2}{\beta^2} f'(\sigma)^2 \Bigg) \right]
\end{equation}
The advantage of this symmetric representation is that we can interpret the operator insertion as the one that creates the FZZT states described in \eqref{eq.SchwarzCoherent} and the RHS of the figure \ref{fig:SchwarzReduction}.\footnote{In this particular case, the eigenstate in \eqref{eq.SchwarzCoherent} is the vacuum state.}
The Jacobian, $\textrm{Pf}(\omega)$, due to the field redefinition from $\varphi,\pi\to f$ is absorbed in the function integral measure $\frac{\mathscrsfs Df}G$ as was shown in \cite{Stanford:2017thb}. Here $\omega$ is the Alekseev-Shatasvili symplectic form given by, \cite{Alekseev:1988ce,Dorn:2006ys,Dorn:2008sw}
\begin{equation}\label{eq.ASmeasure}
	\omega = \delta T_0\,\wedge\int\limits_{-\beta/2}^{\beta/2}\!\!\!d\sigma f'(\sigma)\,\delta f(\sigma) + T_0 \int\limits_{-\beta/2}^{\beta/2}\!\!\!d\sigma\, \delta f'(\sigma)\wedge\delta f(\sigma) + \frac14 \int\limits_{-\beta/2}^{\beta/2}\!\!\!d\sigma \,\frac{f''(\sigma)\wedge\delta f'(\sigma)}{f'(\sigma)^2}~.
\end{equation}
Note that while the Schwarzian action in \eqref{eq.pathIntegralFinal} has an $\mathbb {SL}(2,\mathbb R)$ symmetry, this is further broken to $U(1)$ due to the presence of the operator insertion at $\sigma=\beta/2$. Consequently $G = \mathbb {SL}(2,\mathbb R)$ or $U(1)$ depending on which orbit we are integrating over.
This also suggests that this operator insertion in \eqref{eq.pathIntegralFinal} due to the boundary condition at $\sigma = \beta/2$ can alternatively be understood as integration over circle diffeomorphisms $f$ with nontrivial monodromy specified by $r$ via the trace relation \eqref{eq.MonodromyTrace}. 

\subsection{Boundary conditions as States of the theory}\label{sec.stateInterpretation}
Up to this point we presented the ZZ and FZZT branes in terms of boundary conditions for Liouville theory on an interval. In fact, for what is to come below, a more natural way to think about them is in terms of (boundary) states. Within Liouville theory, transitioning between boundary conditions and states is equivalent to transitioning between an open-string and a closed-string perspective.
Liouville theory is, in fact, one of the most well understood examples of such an open/closed string duality, \cite{Zamolodchikov:2001ah, Nakayama:2004vk}. We study the boundary Liouville theory given by the path integral \eqref{eq.boundaryLiouville} at finite temperature, $T=1/a$.\footnote{Note that this is the temperature of the 2D Liouville theory. It is not the same as $1/\beta$ which will emerge as the natural periodicity of the Euclidean time direction of the Schwarzian theory in the end (if the latter is studied in a thermal ensemble).} In general, one can choose arbitrary conformal boundary conditions at the ends of the strip i.e. in the open-string perspective. The case of Dirichlet boundary condition at both ends, also known as $(1,1)$-Zamolodchikov-Zamolodchikov (ZZ) branes was studied in the context of the Schwarzian theory in \cite{Mertens:2017mtv, Mertens:2018fds}. The generic partition function, $Z_{\mathfrak s,\mathfrak s'}\left[i\frac{a}\beta\right]$, between two different branes is given by, \cite{Zamolodchikov:2001ah, Nakayama:2004vk},
\begin{equation}\begin{aligned}\label{eq.ZZFZZTPartitionFunction}
	Z_{\mathfrak s,\mathfrak s'}\left[i\frac{a}\beta\right] &= \int \!\!\!dP\, \Psi_{\mathfrak s}(P) \Psi_{\mathfrak s'}(-P) \, \chi_P(\tilde q)\,,\\
	\text{where  each of } \mathfrak s \text{ or } \mathfrak{s}' &= \left\{\begin{matrix}(m,n)\in(\mathbb Z,\mathbb Z) \text{ for a generalized $(m,n)$-ZZ brane}\\s\in\mathbb R \text{ for FZZT brane}\end{matrix}\right.~.
\end{aligned}\end{equation}
Here, the modular parameter is $(i\,a/\beta)$ since the size of our open string is $\beta/2$ and $q=\exp\left[ -2\pi \times \frac{a}\beta \right], \ \tilde q = \exp\left[ -2\pi \times \frac\beta{a} \right]$. The particular case of our interest presently is the case where $\mathfrak s= s$ for a FZZT brane and $\mathfrak s'=(1,1)$ for a `basic' ZZ brane. Finally, the Virasoro character corresponding to a non-degenerate state labelled by $P$ is,
\begin{equation}
	\chi_P(q) = \frac{ q^{P^2} }{ \eta\!\left( q\right) }, \quad \eta\!\left( q\right) = q^{1/24} \prod_{n=1}^\infty (1-q^n)~.
\end{equation}
While the states given by the wavefunctions $\Psi_{\mathfrak s}(P)$ can be written in terms of the Ishibashi states, \cite{Ishibashi:1988kg, Onogi:1988qk},\footnote{Our Ishibashi states are normalized as, $\llangle P|q^{L_0/2}|P'\rrangle = \delta(P-P') \, \chi_P(q)$}
\begin{align}
	\langle \mathfrak s| &= \int \!\! dP \Psi_{\mathfrak s}(P) \llangle P|,\label{eq.bdyState}\\
	&\llangle P | = \left\langle \nu_P\right\vert  \left( 1+\frac{L_1\bar L_1 }{ 2\Delta_P } +\cdots\right)~.\label{eq.IshiState}
\end{align}
Here, $\left\vert  \nu_P\right\rangle$ is the state created by the vertex operator $e^{\left( Q+2iP \right)\frac{\varphi}{2b}}$ under the state operator correspondence, with the conformal weight, $\Delta_P = \frac{Q^2}4+P^2$.
Thus the open string partition function between two branes, written as the integral appearing in \eqref{eq.ZZFZZTPartitionFunction},  can equivalently be written as an amplitude between two such closed string states,
\begin{figure}[tbp]
\begin{center}
\includegraphics[width=0.8\textwidth]{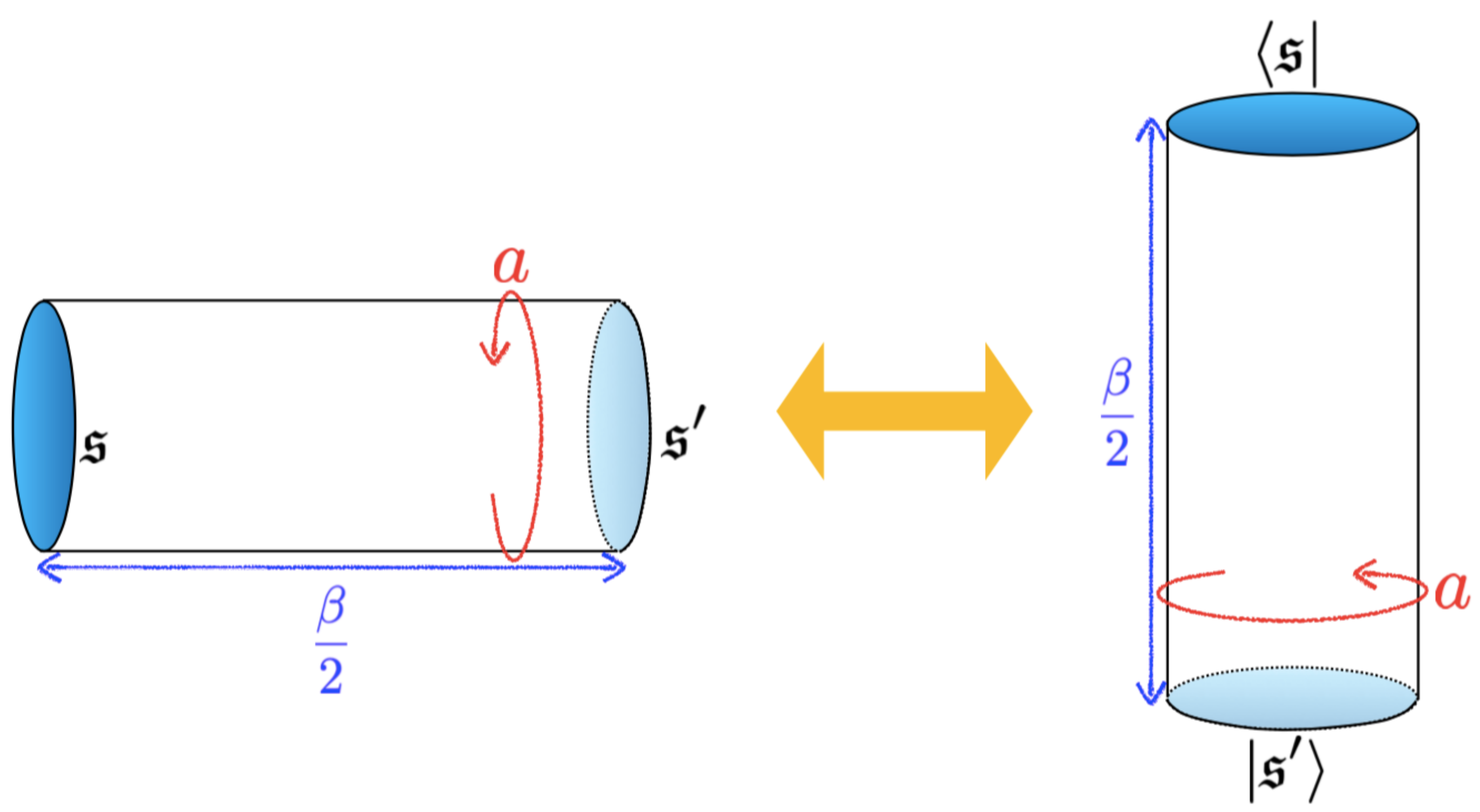}
\caption{An open string Partition function with generic boundary conditions is the same as a closed string amplitude between corresponding boundary states.}
\label{fig.duality}
\end{center}
\end{figure}

\begin{equation}
	Z_{\mathfrak s,\mathfrak s'} [q]= \langle \mathfrak s|\tilde q^{L_0/2}| \mathfrak s'\rangle~.
\end{equation}
The states of importance in this work are,
\be\label{eq.ZZFZZTwavefunctions}\begin{aligned}
	\Psi_{\rm ZZ}(P) &:= \Psi_{(1,1)}(P) = 2^{3/4}2i\pi P \, {\left(\frac{\gamma(b^2)}{8b^2}\right)^{-\frac{iP}b}}\, \frac1{\Gamma\left(1-2iPb\right)\Gamma\left(1-2\frac{iP}b\right)}\\
	\Psi_{\FZZT}^{(s)}(P) &:= \Psi_{s}(P) = e^{2i\pi P s} \left[-\frac{i}{2\pi P} \left(\frac{\gamma(b^2)}{8b^2}\right)^{-\frac{iP}b} \Gamma\left(1+2iPb\right)\Gamma\left(1+2\frac{iP}b\right)\right]~.
\end{aligned}\ee
Here, for the sake of brevity we have used the standard definition $\gamma(x) = \Gamma(x)/\Gamma(1-x)$. This duality needs to be generalized for the case with operator insertions to be useful for our computations.
Unnormalized thermal correlation functions can likewise be computed using the open/closed string identity:
\begin{equation}\label{eq.ZZFZZScatteringAmplitude}
	\textrm{Tr}_{\mathfrak s,\mathfrak s'}\left[e^{-a H} \mathbb V_n(w_n,\bar w_n)\cdots \mathbb V_1(w_1,\bar w_1)\right] = \langle \mathfrak s| \tilde q^{\frac{L_0}2} e^{\frac{2\pi}a \sigma_n L_0 } \mathbb V_n \cdots \mathbb V_2 e^{-\frac{2\pi}a \sigma_{21} L_0 } \mathbb V_1 e^{-\frac{2\pi}a \sigma_1 L_0 } |\mathfrak s'\rangle
\end{equation}
where, we have used following definitions,
\begin{equation}\begin{aligned}
	\tilde q^{L_0} &= \exp\left(-2\pi \frac\beta{a} L_0\right) = \exp(-\beta H) \\
	q^{L_0} &= \exp\left(-2\pi \frac{a}\beta L_0\right) = \exp(-a H) \\
	w_i &= \tau_i + i \sigma_i, \quad \bar w_i = \tau_i - i \sigma_i\\
	&\qquad \sigma_{ij}=\sigma_i-\sigma_j
\end{aligned}\end{equation}
and $\mathbb V_i$ are the vertex operators of the Liouville theory. We have also used $\tau_i = 0 \Rightarrow w_i = -\bar w_i$. This choice has been made with eventual dimensional reduction, $a\to0$, in mind.

Finally putting the results from both these perspective together, in the $b\to0$ limit, we can equate \eqref{eq.pathIntegralFinal} to \eqref{eq.ZZFZZScatteringAmplitude},
\begin{equation}\begin{aligned}\label{eq.ZZFZZSchToState}
	\int\!\!\frac{\mathscrsfs Df}{G} \Big(\cdot\Big) \exp\Big( S[f]+S_N \Big) \sim \langle s| e^{-\frac{\beta L_0}2} \mathbb V_n(\sigma_n,-\sigma_n) \cdots \mathbb V_2(\sigma_2,-\sigma_2) \mathbb V_1(\sigma_1,-\sigma_1) | (1,1)\rangle
\end{aligned}\end{equation}
where, $S[f], S_N$ are the action introduced in \eqref{eq.SchwarzianAfterDoubling} and \eqref{eq.densityOp}; and, $(\cdot)$ corresponds to the insertion of the bilocal operators given by \eqref{eq.OpBiLoc}, $\mathcal O_i(\sigma_i,-\sigma_i)$, in the Schwarzian theory. Also the parameter $r$ on the LHS is related to the parameter $s$ of the FZZT state by,
\begin{equation}\label{eq.relationBWparas}
	\cosh^2(\pi b s) = \frac{r^2}{2\pi b^2} \sin(\pi b^2) = \cos^2(\pi\theta)\, \frac{\sin(\pi b^2)}{\pi b^2}
\end{equation}
For the case of the unperturbed (`standard') Schwarzian theory, we have,
\begin{equation}\begin{aligned}\label{eq.ZZZZSchToState}
	\int\!\!\frac{\mathscrsfs Df}{G} \Big(\cdot\Big) \exp\Big(S[f] \Big) \sim \langle (1,1)| e^{-\frac{\beta L_0}2} \mathbb V_n(\sigma_n,-\sigma_n) \cdots \mathbb V_2(\sigma_2,-\sigma_2) \mathbb V_1(\sigma_1,-\sigma_1) | (1,1)\rangle
\end{aligned}\end{equation}
The tilde in the equations above emphasizes the fact that on both sides of these equations we are computing unnormalized correlation functions. Since we are interested in time-dependent behavior of these correlation functions we will not be careful about the overall time-independent normalizations. Note that in the large central charge limit, $c\to\infty\equiv b\to0$, the Ishibashi states in  \eqref{eq.IshiState}, $\llangle P|\to\left\langle\nu_P\right\vert $, so that in our case they can always be replaced simply with the primary states, which simplifies our task considerably.
Also, $\theta=\pm1 \Leftrightarrow r^2 = 2 \Leftrightarrow s=\pm i$ corresponds to $(1,1)$-ZZ state. This is consistent with our previous observation that $\theta=\pm1$ corresponds to the standard Schwarzian action in \eqref{eq.2DSchwarzian}. In all the subsequent sections, we work only with ZZ-branes of the type $(1,1)$ and therefore we refer to them as $|\ZZ\rangle$ instead of $|(1,1)\rangle$. In our notations $|\FZZT\rangle$ denotes a general FZZT brane and the parameter $s$ is suppressed.

This concludes our introduction relating the 1-dimensional Schwarzian theory to boundary Liouville theory in 2-dimensions. We also related the boundary states of the Liouville theory with different states of the Schwarzian theory. We will now use this setup to compute the correlation functions of the 1-dimensional theory with an aim to unravel their thermal behaviour. Before we proceed to do the exact computations in \autoref{sec.ExactResults}, we derive some of these results using semiclassical analysis in \autoref{sec.Semiclassics}.

\section{Semiclassical results}\label{sec.Semiclassics}
In this section we make use of the natural symplectic structure on the coadjoint orbits of Virasoro to extract the scrambling exponent in pure states. The logic is as follows: first we establish the effective temperature (the `ETH temperature') we should assign to our pure states, including the actual eigenstates of the Schwarzian model. This can be done by studying the one-point functions of bilocal operators ${\cal O}_\ell$. We then go on to the corresponding two-point function of ${\cal O}_\ell's$ and determine the Lyapunov exponent whenever appropriate. In all cases where the one-point function is thermal we find the maximally allowed Lyapunov exponent predicted by taking the ETH temperature seriously. On the other hand, we also establish a phase transition between this ergodic behavior and a non-ergodic phase in which the one-point functions are not periodic in Euclidean time and the corresponding OTOC does not exhibit scrambling behavior. The resulting phase diagram is summarized in Figures \ref{fig:phaseDiag}.
\subsection{One-point function and effective temperature}
We start our exploration of the semi-classical correlators of the model  \eqref{eq.Schwarz} with the simplest example, namely the one-point function of the bilocal operator
\be
{\cal O}_\ell (\sigma_1,\sigma_2) := e^{2\ell\varphi(\sigma_1,-\sigma_2)} 
\ee
where we have emphasized that the operators in the Schwarzian theory descend from the zero-mode of the primary operator of weight $\ell$ in the Liouville picture. This will turn out to be convenient for our next step, when we exploit this connection in order to construct the saddle-point solutions of the Schwarzian path integral with ZZ and FZZT boundary conditions. Later, in section \ref{sec.ExactResults}, we will recover these expressions from a semi-classical expansion of the exact answer for the correlation functions.
\subsubsection*{ZZ branes: Dirichlet condition and energy eigenstates in the Schwarzian}
Here we are interested in the semi-classical answer for the correlation function
\be
\langle E |{\cal O}_\ell (\sigma_1,\sigma_2)| E\rangle~,
\ee
where $|E\rangle$ is an eigenstate of energy $E$. Note that the energy can be defined as the expectation value of the Schwarzian `operator' in the energy eigenstate, \cite{Bagrets:2016cdf, Bagrets:2017pwq, Mertens:2017mtv, Mertens:2018fds}:
$$E:= \langle E| \{F(\sigma),\sigma\}|E\rangle~,$$
with the background temperature, $1/\beta = 0$. Semiclassical limit of this equation corresponds to,\footnote{The appropriate scaling of the energy in this equation is to ensure that we are working with a high-energy state in the semiclassical limit.}
\be
	\{F(\sigma);\sigma\} = \frac {4\pi^2E}C~,
\ee
and, the saddle point solution is given by,
\be\label{eq.SchwarzSaddle}
	F(\sigma) = \tan\left(\pi\sqrt{\frac{2E}C} \sigma \right)
\ee
Recall, in the zero temperature limit the parabolic orbit, ${\rm Diff}(S^1)/SL(2,\mathbb R)$, that is relevant for this case is characterized by the following operator (see also \eqref{eq.OpBiLoc}),
\begin{equation}
	{\cal O}^1_\ell (\sigma_1,\sigma_2) = \left(\frac{F'(\sigma_1)F'(\sigma_2)}{\left(  F(\sigma_1) - F(\sigma_2)   \right)^2}\right)^{2\ell}\,.
\end{equation}
The reason for the superscript `$1$' in the above equations will become clear once we understand the analogous solutions corresponding to Neumann boundary conditions (FZZT branes). The path integral expression that we are interested in involves an integration over the $F(\sigma)$ modes,\footnote{In the $\beta\to\infty$ limit, the fields $F(\sigma)=f(\sigma)$}
\be
\int \frac{\mathscrsfs{D}F}{{\rm SL}(2,\mathbb{R})}{\cal O}^1_\ell (\sigma_1,\sigma_2)  e^{C\int d\sigma  \left\{F(\sigma);\sigma \right\}}\,.
\ee
However, in the semiclassical limit it can be evaluated at the above saddle point solution, \eqref{eq.SchwarzSaddle},
\begin{equation}\label{eq.SchwarzSaddleExp}
	\langle E| {\cal O}^1_\ell (\sigma_1,\sigma_2) |E\rangle \approx \left(\frac{\pi^2}{\beta_E^2}\frac1{\sin^2\left(\frac\pi{\beta_E} \sigma_{12}\right) }\right)^{2\ell}~, \qquad 1/\beta_E = T_E = \sqrt{\frac{2E}C}
\end{equation}
This is the same as the thermal correlation function at an effective temperature, $T_E$. Later in \autoref{sec.ExactResults} we obtain the same result using the exact methods described in the previous section.

\subsubsection*{FZZT branes: Neumann condition and Schwarzian coherent states}
We are now interested in the semi-classical answer for the correlation function
\be
\langle E_r |{\cal O}_\ell (\sigma_1,\sigma_2) | E_r\rangle
\ee
where, $|E_r\rangle$ corresponds to the local operator discussed in \eqref{eq.SchwarzCoherent}. This insertion restricts us to the orbit Diff($S^1$)/U$(1)$, as was deduced in the previous section and corresponds to the zero mode of a classical solution to Liouville theory with a Dirichlet boundary condition at $\sigma = 0$ and a Neumann boundary condition at $\sigma = \beta/2$. In this case the classical solution \cite{Dorn:2006ys} can be written as
\be
{\cal O}^\theta_\ell (\sigma_1,\sigma_2) = \left(\frac{\pi^2\theta^2}{\beta_E^2} \frac{f'(\sigma_1)f'(\sigma_2)}{\sin^2\left( \frac{\theta \pi}{\beta_E}\left(  f(\sigma_1) - f(\sigma_2)   \right)\right)}\right)^{2\ell}
\ee
The superscript $\theta$ parametrizes the family of Neumann boundary conditions corresponding to our coherent states. We note that  setting $\theta = 1$ recovers the pure Dirichlet case, explaining the choice of superscript above. Also note that the background temperature for this orbit is taken to be the same as the effective temperature $T_E=1/\beta_E$ induced by external states $|E\rangle$.

 In order to get the Schwarzian correlation function we still need to integrate over the circle diffeomorphism $f$. We must thus evaluate the path integral
\be
\int \frac{\mathscrsfs{D}f}{U(1)}{\cal O}^\theta_\ell (\sigma_1,\sigma_2) e^{-\frac{ 4\pi C \, \theta}{\beta} \, \frac{f'(\sigma_3)}{\tan\left(\pi\,\theta\right)} -C \int \!\!d\sigma \{f(\sigma);\sigma\} }\,.
\ee
where we have explicitly written the path integral in terms of the $f$ fields as opposed to $F$ fields. For $C\gg 1$, this expression is again evaluated via saddle point resulting in the same solution as above $f(\sigma) = \sigma$, and thus in the matrix element
\be\label{eq.ZZFZZTsemicalssical}
\langle E_{r} |{\cal O}_\ell (\sigma_1,\sigma_2) | E_{r}\rangle =\left[\left(\frac{\pi}{\beta_\theta}\right)^2 \frac{1}{\sin^2\left( \frac{\pi}{\beta_\theta}  \sigma_{12}\right)}\right]^{2\ell}
\ee
with effective temperature
\be \label{eq.EffTempCoherent}
T_\theta=1/\beta_\theta = \theta \sqrt{\frac{2E}C} \,.
\ee
The reader will recall that solutions on this orbit are classified by their monodromy, whence $\theta \in \mathbb{R}$ corresponds to the elliptic class of solutions while $\theta = ip \in i\mathbb{R}$ corresponds to the hyperbolic class. The time parameter $\sigma$ appearing in the correlation function has the interpretation of Euclidean time in the 1D theory, so that the result is thermal for $\theta \in \mathbb{R}$ and non-thermal for $\theta \in i\mathbb{R}$. We note that this is in perfect agreement with the results of \cite{Jensen:2019cmr,David:2019bmi} who consider the scrambling behavior of the 2D identity block and finds an effective temperature analogous to \eqref{eq.EffTempCoherent} also in the elliptic case (i.e. for operators below the BTZ threshold). It is interesting to note that \cite{Anninos:2018svg}, who study properties of de Sitter horizons, also find the oscillatory to exponential cross-over of the OTOC, which they associate with the different Virasoro coadjoint orbits. A soft-mode action of the Schwarzian type appears in their work as the boundary action of AdS$_2$, which has been glued to a dS$_2$ region in the IR.

We would now like to go on and calculate the semi-classical two-point function of bilocals, which will allow us to extract the OTOC in the pure states $|\Psi\rangle$. We will arrive at this result by using the symplectic structure of Alekseev and Shatashvili \cite{Alekseev:1988ce,Dorn:2006ys}, allowing us to compute the semi-classical expectation value of commutators of the relevant operators.
\subsection{Semiclassical OTOC and chaos conjecture}\label{sec.SemiclassicalChaos}

As discussed in \cite{Dorn:2006ys,Dorn:2008sw}, the standard symplectic form $\omega = \int d\pi \wedge d \varphi$ subject to the boundary conditions considered in section \ref{sec.BoundaryLiouville} gives rise to the Poisson bracket
\bea
\left\{ f(\sigma_1), f(\sigma_2) \right\}_{\rm PB} &=& \frac{1}{4T_0}\left( \frac{\sinh\left( 2\sqrt{T_0}\lambda(\sigma_1,\sigma_2)\right)}{\sinh\left(  2\pi\sqrt{T_0}\right)}    - \frac{\lambda(\sigma_1,\sigma_2)}{\pi}\right)\,,\\
 \lambda(\sigma_1,\sigma_2) &=& f(\sigma_1) - f(\sigma_2)  - \pi \epsilon(\sigma_1-\sigma_2)\nonumber
\eea
between two Schwarzian soft modes. Here $\epsilon(x) = 2n+1, \,\, x \in (2\pi n, 2\pi (n+1))$ is the stair step function, which will play no further role in our analysis, while $T_0$ is the constant representative defined in \eqref{eq.TconstantOrbit}. For large separation $\sigma_1-\sigma_2$ and evaluated on the saddle point \eqref{eq.SchwarzSaddle}, this Poisson bracket takes the simple form
\be\label{eq.Poisson}
\left\{ f(\sigma_1), f(\sigma_2) \right\}_{\rm PB}  \sim  \frac{\sinh\left( 2\sqrt{\frac{2E}CT_0}(\sigma_1-\sigma_2)\right)}{4T_0 \sinh\left(  2\pi\sqrt{T_0}\right)}
\ee
Using to the standard Dirac quantization prescription, this then allows us to compute the semi-classical commutator
\be
\left[ f(\sigma_1), f(\sigma_2) \right]_{\rm s.c.} = -i\hbar \left\{ f(\sigma_1), f(\sigma_2) \right\}_{\rm PB}
\ee
where we have of course $\hbar = 1/C$ by comparing to the action \eqref{eq.Schwarz}. We now explain how this allows us to extract the quantum Lyapunov exponent with respect to our pure states $|\Psi\rangle$. The quantum Lyapunov exponent is diagnosed from a correlation function with four time insertions of the type
\be
G^{\rm OTO}_{\ell_1,\ell_2} (t_1,t_2,t_3,t_4) =  \langle \Psi |  {\cal O}_{\ell_1}(t_1,t_2) {\cal O}_{\ell_2}(t_3,t_4)|\Psi \rangle,
 \ee
  where the Lorentzian times are ordered, such that $t_1 < t_3 < t_2 < t_4$. This can be obtained as an analytic continuation from the Euclidean correlation function $\langle \Psi |  {\cal O}(\sigma_1,\sigma_2) {\cal O}(\sigma_3,\sigma_4)|\Psi \rangle$ by defining\footnote{In the thermal case one instead often displaces the Lorentzian times a quarter turn around the thermal circle
  $$ \sigma_1 = -\frac{\beta}{4} + i t_1\,,\quad \sigma_2 = \frac{\beta}{4} + i t_1\,,\quad \sigma_3 =  -\frac{\beta}{4} + i t_2\,,\quad \sigma_4 = \frac{\beta}{4} + i t_2\,.$$
  
  One could consider a similar arrangement by inserting the ETH temperature associated with the pure state, but the above arrangement appears more natural in our context. } 
\be
\sigma_i = it_i + \epsilon_i\,,\qquad \textrm{with}\qquad \epsilon_1 < \epsilon_3 <\epsilon_2 < \epsilon_4
\ee
The resulting correlation function then depends on the precise analytical structure of the correlator as a function of complex time insertions. The problem was solved in \cite{Mertens:2017mtv} for the thermal OTO by appealing to the $R$-matrix of Ponsot and Teschner \cite{Ponsot:1999uf} together with a plausible ansatz about its behavior in the semiclassical limit. In order to find the OTO in eigenstates we take a different route, which also applies to the thermal case, where it agrees with the answer found in \cite{Mertens:2017mtv}. It would be interesting to further understand how these two methods are related.

Thanks to the semi-classical results we established above, in conjunction with the symplectic form \eqref{eq.Poisson}, we can sidestep this somewhat involved procedure. Before we describe this argument let us introduce some notation. For the purposes of being explicit about analytic continuation it is often useful to formally split up a bilocal operator as ${\cal O}_\ell (\sigma_1, \sigma_2) \sim V(\sigma_1) V(\sigma_2)$, and thus denote the corresponding correlation function
\be
\langle {\cal O}_{\ell_1}(\sigma_1,\sigma_2) {\cal O}_{\ell_2}(\sigma_3,\sigma_4) \rangle \quad \Leftrightarrow  \quad \langle V(\sigma_1) V(\sigma_2) W(\sigma_3) W(\sigma_4) \rangle \,.
\ee
This is simply a formal device that makes clear that we may consider any ordering of (Lorentzian) time insertions on all legs, such as the time ordered arrangement $t_1 < t_2 < t_3 < t_4$, or the out-of-time ordered one $t_1 < t_3 < t_ 2< t_3$. The order of arranging the `split' bilocals simply expresses the corresponding time-order of the $2n$ times of a correlation function of $n$ bilocal operators.

\begin{figure}[t]
\begin{center}
\includegraphics[width=0.8\columnwidth]{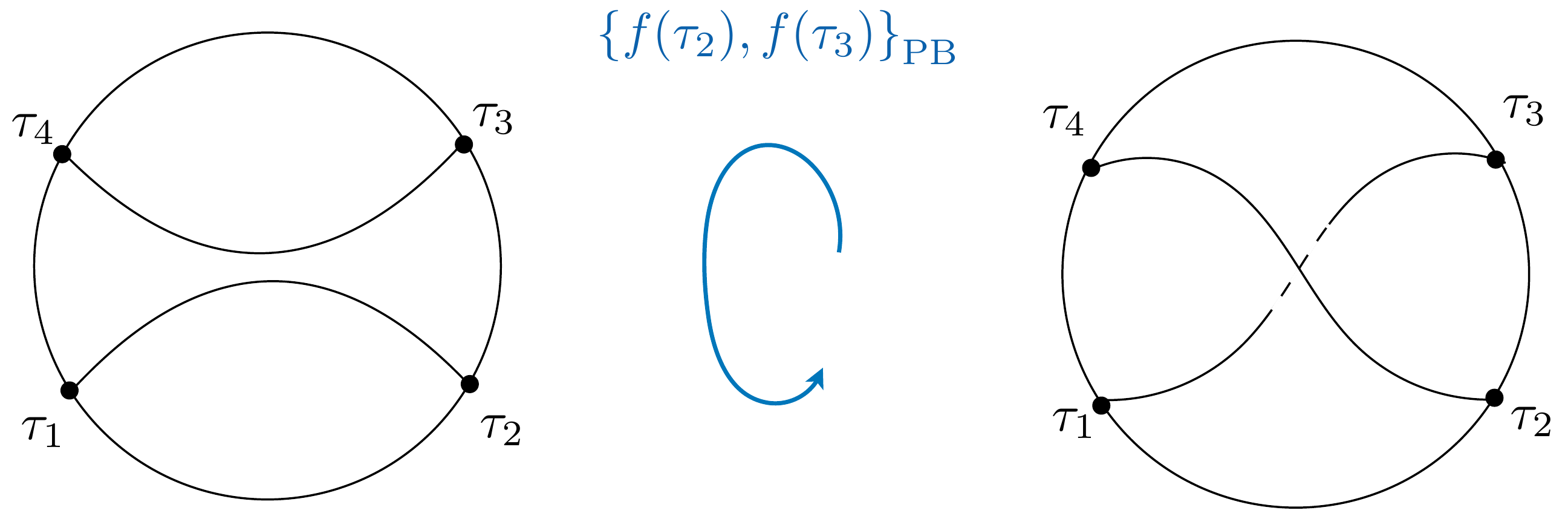}\\
\end{center}
\caption{Illustration of our computation of the semiclassical OTO by saddle point. In order to get the operator in OTO order as shown, we must twist two of the insertions around each other. Quantum mechanically this corresponds to the 'second-sheet' analytic continuation from Euclidean to Lorentzian times and is equivalent to the insertion of a commutator $\left[f(\sigma_1),f(\sigma_2)  \right]$ when writing the operators in terms of integrals over the Goldstone mode $f$. This commutator can be computed semiclassically using the AS symplectic form to generate the Poisson bracket $\left\{f(\sigma_1),f(\sigma_2)  \right\}$ living on the appropriate Virasoro coadjoint orbit (Details in section \ref{sec.SemiclassicalChaos}).}
 \label{fig:4ptTwist}
\end{figure}

With this formal device in place, the difference between a time-ordered and an out-of-time-order correlation function is given by the insertion of a commutator
\be
G_{\ell_1\ell_2}^{\rm TO}(t_1,t_3,t_2,t_4) - G_{\ell_1\ell_2}^{\rm OTO} (t_1,t_2,t_3,t_4)= \langle W(t_1) \left[ W(t_3),V(t_2)\right]  V(t_4)\rangle
\ee
This relation is elementary, but we can also understand it in terms of the analytic continuation from the Euclidean correlator. The difference between the continuation to a time-ordered and an out-of-time order configuration of Lorentzian times is due to crossing a branch cut in the complex $\sigma$ plane. The discontinuity across the cut is, once again, given by the insertion of the commutator in the correlation function. We can view the construction of the OTOC from a TOC as a braiding operation, as illustrated in Figure \ref{fig:4ptTwist}.

In the case at hand, we can evaluate the commutator by using the Poisson bracket \eqref{eq.Poisson} together with the various semiclassical saddle points relevant to our pure states
 \eqref{eq.SchwarzSaddle}. We thus have
\bea
G_{\ell_1\ell_2}^{\rm TO} - G_{\ell_1\ell_2}^{\rm OTO} &=&  \frac{1}{C}\frac{\delta {\cal O}_{\ell_1}}{\delta f(t_2)} \frac{\delta {\cal O}_{\ell_2}}{\delta f(t_3)}\left\{  f(t_2), f(t_3) \right\}_{\rm PB}\Biggr|_{f=f_{\rm saddle}}\nonumber\\
&\sim &  \frac{1}{C} \sinh\left( 2\pi\theta \sqrt{\frac{2E}C} t_{23}  \right)
\eea
The tilde in this expression denotes a time-independent proportionality factor, which includes the normalization of the symplectic form \eqref{eq.Poisson} $1/\sin(\pi\theta)$ in the denominator. This factor diverges at the parabolic orbit $\theta = 1$. We implicitly regulate this UV divergence by introducing an $\epsilon$ factor, which we view as analogous to the divergent prefactor in \cite{Roberts:2014ifa}.
This expression thus tells us that the OTO correlation function behaves maximally chaotically at the appropriate eigenstate temperature 
\be\label{eq.MaximalEigenstateOTOC}
 \langle \Psi | {\cal O}_{\ell_1} (t_1,t_2) {\cal O}_{\ell_1}(t_3,t_4) |\Psi\rangle_{\rm OTO}  \sim 1-\frac{\#}{C} e^{\lambda t_{23}}\,,\qquad \textrm{with}\quad\lambda = 2\pi/\beta_\Psi
 \ee
 where $|\Psi\rangle$ is either $|E(k)\rangle$ or one of the coherent states $|E_r\rangle$ for $r<\sqrt{2}$. The former corresponds to parabolic and the latter to elliptic monodromy of the associated Hill's equation. On the other hand, if we dial the parameter $r$ characterizing the boundary state through $r_c=\sqrt{2}$ we find the non-ergodic OTOC
 \be
  \langle E_r | {\cal O}_{\ell_1} (t_1,t_2) {\cal O}_{\ell_1}(t_3,t_4) |E_r\rangle_{\rm OTO}  \sim 1 - \frac{\#}{C}
  e^{i\lambda t_{23}}\,,\qquad \textrm{with}\quad\lambda = 2\pi\sqrt{\frac{2E}C}p
 \ee
 for $\theta = ip \in i\mathbb{R}$. We have thus uncovered a phase transition in the scrambling behavior of the Schwarzian theory\footnote{And thus also in the IR limit of the SYK model and related many-body systems \cite{Gurau:2010ba,Bonzom:2011zz,Witten:2016iux, Klebanov:2016xxf}}, whereby the model changes from maximally chaotic behavior with exponentially growing OTOCs to a phase that does not scramble at all and the OTOC behaves in an oscillatory fashion. As was remarked in an analogous two-dimensional CFT context, this is an analytical example of a transition between an ergodic and a non-ergodic phase and as such deserves further study.

\subsection{An alternative derivation}\label{sec.AlternativeDerviation}
For the expectation values with respect to the FZZT density operator $\rho= | E_r \rangle \langle E_r|$ we can obtain the chaos exponent directly from a perturbative quantization of the action \eqref{eq.pathIntegralFinal}. Since this discussion parallels that in \cite{Yoon:2019cql,David:2019bmi,Jensen:2019cmr}, who computed the chaos exponent in thermal states excited by a heavy operator, we will be brief. The idea is to expand the action appearing in the exponent of \eqref{eq.pathIntegralFinal} in fluctuations 
\be
f(\sigma) = \tan\left(\frac{\pi}{\beta_{\rm eff}} \left( \sigma + \varepsilon(\sigma) \right) \right)\,,
\ee
where $\beta_{\rm eff}$ is the effective temperature defined in \eqref{eq.CoherentEffectiveTemperature}.
We then compute the Euclidean correlation function of two bilocals $\langle {\rm O}_{\ell_1}(\sigma_1, \sigma_2){\rm O}_{\ell_2}(\sigma_3, \sigma_4) \rangle $ using the propagator of the fluctuation $\langle\varepsilon(\sigma) \varepsilon(\sigma')\rangle$ given, for example, in \cite{Maldacena:2016hyu,Maldacena:2016upp}. The leading perturbative result, continued into the chaos region reads
\begin{align}
\frac{\langle {\rm O}_{\ell_1}(\sigma_1, \sigma_2){\rm O}_{\ell_2}(\sigma_3, \sigma_4) \rangle}{\langle {\rm O}_{\ell_1}(\sigma_1, \sigma_2) \rangle\langle  {\rm O}_{\ell_2}(\sigma_3, \sigma_4)\rangle} &\sim -2\pi\ \frac{ \sin \left(\frac{\pi}{\beta_{\rm eff}} (\sigma_1+\sigma_2-\sigma_3-\sigma_4)\right) +\ldots  }{ \sin \left(\pi\frac{\sigma_{12}}{\beta_{\rm eff}}\right) \sin \left(\pi\frac{\sigma_{34}}{\beta_{\rm eff}}\right) }\\[5pt]
\xrightarrow[\sigma_{3,4}\to \pm\epsilon_{34}]{\sigma_{1,2}\to it\pm\epsilon_{12}}&\quad 1 -\frac{\#}{C }\ \frac{ \sinh \left(\frac{2\pi}{\beta_{\rm eff}} t\right)}{ \epsilon_{12}\epsilon_{34} } +\ldots  \label{eq.ScramblePerturbationTheory}
\end{align}
In the above expressions the ellipsis denotes subleading corrections to exponential behaviour and $\epsilon_{12},\epsilon_{34}$ originate from the continuation procedure, as indicated. We are keeping them in the final result to play the role of UV regulators  \cite{Roberts:2014ifa}. In the second line we have added the leading disconnected contribution which is always present.
From this expression we deduce, again, that $\lambda = \frac{2\pi}{\beta_{\rm eff}}$, i.e. that the coherent states are maximally scrambling at the effective temperature deduced from ETH. Of course we must ensure to be in the ergodic region ($r< \sqrt{2}$) for this result to hold.

 We now describe a computation that exploits the boundary-CFT perspective of \cite{Fateev:2000ik,Zamolodchikov:2001ah} together with \cite{Anous:2019yku} to give yet another derivation of the result above.

\begin{figure}[t]
 \begin{center}
 	\includegraphics[width=0.60\textwidth]{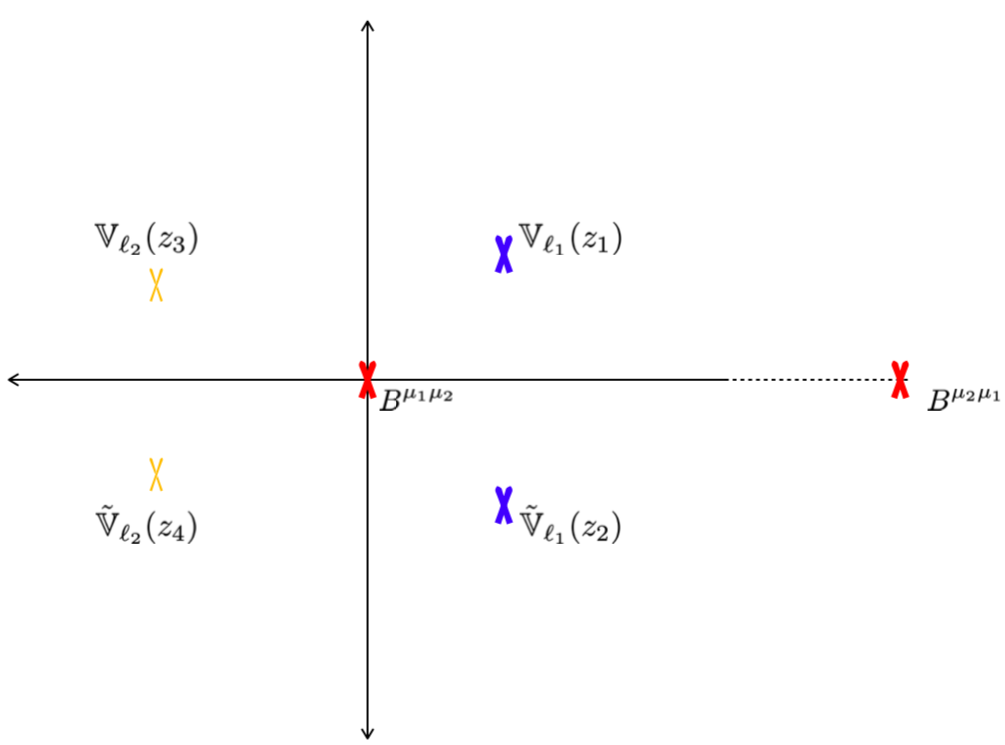}
 \end{center}
 \caption{The configuration of various insertions on the two-dimensional complex plane that evaluates the bilocal two-point functions with ZZ and FZZT boundary conditions. Operators $B$ are those that impose the ZZ and FZZT boundary conditions on positive and negative real axis respectively. While the tilde-operators are the mirror operators after the doubling trick.  }
 \label{fig.insertion-confg}
 \end{figure}

 As pointed out in \cite{Lam:2018pvp}, the dimensionless coupling of the Schwarzian theory is $\sim C/\beta$, and thus the large$-C$ limit is equivalent to a high-temperature limit at fixed $C$.  One may then take the high-temperature limit directly in the Liouville theory before descending to the Schwarzian \cite{Lam:2018pvp}. We thus want to calculate the two-point function of the Liouville vertex operator in the UHP with appropriate boundary conditions on the real axis. Recall that the boundary states $|E_r\rangle$ correspond to ZZ boundary conditions on the positive real axis and FZZT boundary conditions on the negative real axis.  This means that we are really evaluating a six-point function on the plane in the geometry shown in Figure \ref{fig.insertion-confg}.
 There are boundary operators implementing the ZZ and FZZT boundary conditions at the origin and at infinity, as well as the two physical operators in the UHP and their two mirror operators in the LHP. We formally allow the mirror operators' positions to be arbitrary, as indicated in Figure \ref{fig.insertion-confg}. In this way each pair of mirror operators descends to an operator of the type ${\cal O}_{\ell}(\sigma_1,\sigma_2)$ in the Schwarzian limit, notably admitting two arbitrary time insertions ($\sigma_2 \neq - \sigma_1$). It was shown in \cite{Anous:2019yku}, under the assumption of identity dominance, that such a six-point function reduces to an effectively thermal result for the four-point function of the vertex operators alone. In the present context identity dominance is no longer an assumption, as it follows from the OPE coefficients of boundary Liouville theory \cite{Fateev:2000ik,Zamolodchikov:2001ah}. Hence we  evaluate the correlation function by contracting each operator with its mirror operator into the identity, using the boundary OPE derived in \cite{Fateev:2000ik,Zamolodchikov:2001ah} and like-wise contract the boundary operators into the identity with the result% \js{power of prefactor}
\be\label{eq.identityBlock}
\frac{\langle {\rm O}_{\ell_1}(\sigma_1, \sigma_2){\rm O}_{\ell_2}(\sigma_3, \sigma_4) \rangle_r^{\rm TO}}{\langle {\rm O}_{\ell_1}(\sigma_1, \sigma_2) \rangle_r \langle  {\rm O}_{\ell_2}(\sigma_3, \sigma_4)\rangle_r}  ={\rm J}(z) (1-z)^{2\Delta_{\ell_1}} \mathcal{F}_{\rm id} \left[\, {}^{\Delta_{\ell_1}}_{\Delta_{\ell_1}} \,{}^{\Delta_{\ell_2}}_{\Delta_{\ell_2}} ; 1-z\right]\,.
\ee 
Here the conformal block is evaluated on the thermal coordinate \cite{Fitzpatrick:2015zha,Anous:2019yku} $ z_i =e^{i\frac{2\pi}{\beta_{\rm eff}}\sigma_i}$ so that the cross ratio takes the form $
z = \frac{\sin\left(  \frac{\pi}{\beta_{\rm eff}}\sigma_{13}   \right) \sin\left(  \frac{\pi}{\beta_{\rm eff}}\sigma_{24}   \right)}{\sin\left(  \frac{\pi}{\beta_{\rm eff}}\sigma_{23}   \right)  \sin\left(  \frac{\pi}{\beta_{\rm eff}}\sigma_{14}   \right)}
$ and ${\rm J}$ is the Jacobian for the coordinate transformation $\sigma_i \rightarrow z_i$ \cite{Anous:2019yku} whose specific form plays no further role in this analysis. Contracting each operator with its mirror operator means that $z\rightarrow 1$. Under the same analytic continuation as above we obtain
\be\label{eq.CrossRatio}
1-z\quad\xrightarrow[\sigma_{3,4}\to \pm\epsilon_{34}]{\sigma_{1,2}\to it\pm\epsilon_{12}} \quad -\frac{\pi^2\epsilon_{12}\epsilon_{34}}{\beta_{\rm eff}^2}e^{-\frac{2\pi}{\beta_{\rm eff}}t}
\ee
Furthermore, in order to move from the Euclidean configuration of the $\sigma_i$ to the Lorentzian OTO configuration imposed by the $\epsilon_i$, we moved the cross ratio around the branch point at $z=1$, so that now the block is evaluated on the second sheet \cite{Roberts:2014ifa,Anous:2019yku}. 
In the Schwarzian limit, this gives again, to leading order
\be
\frac{\langle {\rm O}_{\ell_1}(\sigma_1, \sigma_2){\rm O}_{\ell_2}(\sigma_3, \sigma_4) \rangle_r^{\rm OTO}}{\langle {\rm O}_{\ell_1}(\sigma_1, \sigma_2) \rangle_r \langle  {\rm O}_{\ell_2}(\sigma_3, \sigma_4)\rangle_r}  = 1 - \frac{\#}{C\epsilon_{12}\epsilon_{34}}e^{\frac{2\pi}{\beta_{\rm eff}}t} + \cdots
\ee 
 Thus once again we find the maximally chaotic behavior \eqref{eq.MaximalEigenstateOTOC} of the OTOC in pure states. As in \eqref{eq.ScramblePerturbationTheory} we should keep in mind that the exponential behavior is true so long as we stay in the range $r< \sqrt{2}$. For $r>\sqrt{2}$ the effective temperature becomes imaginary, and the OTOC oscillates.

The recent papers \cite{David:2019bmi,Jensen:2019cmr} consider a closely related\footnote{In fact the effective-field-theory of the identity block \cite{Cotler:2018zff,Haehl:2018izb} employed in \cite{Jensen:2019cmr} makes the analogy even closer, taking the form of a Schwarzian-like description for this object.} computation in the context of two-dimensional CFT: they insert two heavy operators at spatial infinity into an already thermal system. The resulting configuration is then again approximately thermal, but at the modified temperature $\beta_{\rm eff}$, depending on the heavy state insertions. The scrambling exponent found in \cite{David:2019bmi,Jensen:2019cmr} takes on the value $2\pi/\beta_{\rm eff}$ with respect to this effective temperature, in accordance with the results we find for the Schwarzian theory.

\section{Exact results}\label{sec.ExactResults}
In this section we describe how to obtain the exact results for the Schwarzian correlation functions we discussed above semi-classically. We will also show that these expression result from a suitable semi-classical expansion of the full answers, giving a second independent derivation of our results on eigenstates and the behavior of correlation functions therein. Our main technical vehicle to obtain full exact expressions for correlation functions is the descent construction described in section \ref{sec.BoundaryLiouville}, in other words we will find the Schwarzian answers as limits of those obtained in boundary Liouville theory using the identities \eqref{eq.ZZFZZSchToState} and \eqref{eq.ZZZZSchToState}.

\subsection{Computations between ZZ branes}\label{sec.exactZZ}
We now use the boundary state perspective to find the one and two-point functions of bilocal operators in the presence of Dirichlet boundary conditions. This will allow us to confirm our semiclassical expectations above, as well as to show the maximal eigenstate chaos conjecture once more from a different perspective. One and two-point functions with Dirichlet conditions have previously been computed by \cite{Mertens:2017mtv,Lam:2018pvp}. We repeat these calculations for two reasons: firstly to extend those results to include extended ETH as well as the chaos exponent in eigenstates; and secondly, to express everything in our  own conventions \footnote{A caveat for the reader interested in reproducing the detailed calculations: our conventions are fully aligned with \cite{Zamolodchikov:2001ah}, but differ  in some places from the ones used in \cite{Mertens:2017mtv,Lam:2018pvp}.} before moving on to the general case including FZZT states.
%firstly we demonstrate ETH in presence of heavy operators and extended ETH for the computation of the chaos exponent. Secondly, we use this opportunity to express everything in our  own conventions

\subsubsection{Bilocal one-point function \texorpdfstring{$\langle \ZZ|\mathbb V_\ell(x,\bar x)|\ZZ\rangle$}{<ZZ|V-l|ZZ>}}\label{sec.BilocalOnePointZZ}
We are interested in evaluating a particular case of \eqref{eq.ZZZZSchToState} with only one operator insertion. Recall, the ZZ-wavefunction can be written as, \eqref{eq.ZZFZZTwavefunctions},
\be
\Psi_{\rm ZZ}(P) := \langle {\rm ZZ}|P\rangle = 2^{3/4}2i\pi P \, {\left(\frac{\gamma(b^2)}{8b^2}\right)^{-\frac{iP}b}}\, \frac1{\Gamma\left(1-2iPb\right)\Gamma\left(1-2\frac{iP}b\right)}\,.
\ee
 Putting together the ingredients we assembled in \autoref{sec.stateInterpretation}, the expression for the expectation value $\langle {\rm ZZ}|\mathbb V_\ell(z,\bar z)|{\rm ZZ}\rangle$ reads in full detail
\begin{eqnarray}\label{eq.ZZOnePointFunction}
	2^{3/2} \,\pi^2\int\!\! dP^2 \,dR^2 \, \frac{ \left(\frac{\gamma(b^2)}{8b^2}\right)^{i\frac{P-R}b} }{ \Gamma\left(1+2ibP\right) \Gamma\left(1+2i\frac Pb\right)\Gamma\left(1-2ibR\right) \Gamma\left(1-2i\frac Rb\right) } \left\langle\nu_R\right\vert \mathbb V_\ell(z,\bar z)\left\vert \nu_P\right\rangle
\end{eqnarray}
Let us now start simplifying this expression, starting with the matrix element of the vertex operator between primary states,
\begin{equation}\begin{aligned}
	\left\langle\nu_R\right\vert \mathbb V_\ell(x,\bar x)\left\vert \nu_P\right\rangle &=\langle \nu_R| e^{-\frac{4\pi}a \left(\frac\beta2-\sigma\right) L_0 } \mathbb V_\ell e^{-\frac{4\pi}a \sigma L_0 } |\nu_P\rangle\\
	&= e^{-\frac{2\beta\pi} a\Delta_R} e^{- \frac{4\pi}a \sigma \left(\Delta_P-\Delta_R\right)} \left\langle\nu_R\right\vert \mathbb V_\ell(0,0) \left\vert \nu_P\right\rangle\\
	&= e^{-\frac{2\beta\pi} a\Delta_R} e^{- \frac{4\pi}a \sigma \left(\Delta_P-\Delta_R\right)} \,\frac1{2b}\int \!\!d\varphi \left\langle\nu_R\right\vert \left.\varphi\rangle \langle\varphi\right.\left\vert \nu_P\right\rangle e^{2\ell\varphi}
\end{aligned}\end{equation}
where, $w,\bar w = \tau\pm i\sigma$ are the coordinates on the open string; and, $x,\bar x = - i \tau\pm \sigma = -i w,-i\bar w$ are the coordinates on the dual closed string. We have also used the fact that we are working with a chiral-CFT with only the holomorphic sector. The projection of the wavefunctions on the $\varphi$-basis is given by \cite{Zamolodchikov:2001ah},
\begin{equation}\begin{aligned}
	\left\langle\nu_R\right\vert \left.\varphi\right\rangle &= 2\frac{\left(8b^4\right)^{-i\frac Rb}}{\Gamma\left(2i\frac Rb\right)}\, K_{-2i\frac Rb}\left(\frac{e^{\varphi/2}}{\sqrt2b^2}\right)\,,\\
	\left\langle\varphi\right.\left\vert \nu_P\right\rangle &= 2\frac{\left(8b^4\right)^{i\frac Pb}}{\Gamma\left(-2i\frac Pb\right)}\, K_{2i\frac Pb}\left(\frac{e^{\varphi/2}}{\sqrt2b^2}\right) \,.
\end{aligned}\end{equation}
Thus the matrix element boils down to
\begin{equation}\label{eq.dozz}
	\left\langle\nu_R\right\vert \mathbb V_\ell(x,\bar x)\left\vert \nu_P\right\rangle = \frac1{2b} \left({8b^4}\right)^{2\ell+i\frac{P-R}b} \frac{ \Gamma\left(2\ell\pm\frac ib(P\pm R)\right) }{ \Gamma\left(-2i\frac Pb\right)\Gamma\left(2i\frac Rb\right) \Gamma(4\ell)}\, e^{-\frac{2\beta\pi} a\Delta_R} e^{- \frac{4\pi}a \sigma \left(\Delta_P-\Delta_R\right)} ~.
\end{equation}
In view of our descent to the 1D Schwarzian theory, we need to additionally make the change of variables, $P=b\, \mathfrak p, R=b\, \mathfrak r$, since, from this point of view, we are interested in the $b\to0$ limit. Recall that the operator insertions in Liouville theory corresponding to the bilocal operator insertions in the 1D theory are of the form,
\be
	\mathbb V_\ell(x,\bar x) = e^{2\ell \varphi} \quad \underrightarrow{\text{under 1D reduction}} \quad\left( \frac{\sqrt{f'(\sigma)f'(-\sigma)}}{|f(\sigma)-f(-\sigma)|} \right)^{2\ell}
\ee
For the most part we are interested in operators satisfying $\ell\sim\mathcal O(1)$, in which case the conformal dimension of the 2D vertex operator is $\Delta_\ell = 2b\ell\left(Q-2b\ell\right)$. Since we want the boundary states to be of similar energies as the insertions themselves, $\Delta_P\sim \Delta_\ell \Rightarrow P\sim\pm\frac Q2\mp2b\ell$.\footnote{More accurately, we want to insert a complete set of states for the 1-dimensional theory, which is equivalent to insertion of ZZ branes with energies scaled as described here. On the other hand, as we will see in the next section, the insertion of FZZT states is equivalent to choosing a `heavy' state.} Up to the constant shift, $\frac Q2$, which will not matter, we get $P\sim \mathcal O(b)$. Now we reinsert the expression for the matrix element into the full expectation value \eqref{eq.ZZOnePointFunction}. In anticipation of the $b\rightarrow 0$ limit to come, we already restrict the Ishibashi states to their leading term (the primary state) and write
\begin{align}
	\langle \ZZ|\mathbb V_\ell(x,\bar x)|\ZZ\rangle &= 2^{1/2}b^3\, \left({8b^4}\right)^{2\ell} \int\!\! d\mathfrak p^2 \,d\mathfrak r^2 \,\sinh(2\pi \mathfrak p)\sinh(2\pi\mathfrak r) \, e^{-\frac{2\beta\pi} a \left(\frac{Q^2}4+b^2\mathfrak r^2\right)} e^{\frac{4\pi}a \sigma\, b^2 \left(\mathfrak r^2 -\mathfrak p^2\right)} \nonumber\\
	& \hspace{3cm} \times \frac{ \Gamma\left(2\ell\pm i(\mathfrak p\pm \mathfrak r)\right) }{ \Gamma(4\ell)} \, \left\{\frac{ \left(b^2\gamma(b^2)\right)^{i\left(\mathfrak p-\mathfrak r\right)} }{ \Gamma\left(1+2ib^2\mathfrak p\right) \Gamma\left(1-2ib^2\mathfrak r\right) } \right\}\,. \nonumber
\end{align}
We can now straightforwardly take the limit $b\rightarrow 0$ of the above expression, giving the final result for the expectation value in the form,
\begin{align}
	\langle \ZZ|\mathbb V_\ell(x,\bar x)|\ZZ\rangle &=2^{1/2}b^3 \left({8b^4}\right)^{2\ell}e^{-\frac{\pi \beta Q^2}{ 4a }} \!\!\!\int\!\!\! d\mathfrak p^2 d\mathfrak r^2 \, e^{2\pi(\mathfrak p+\mathfrak r)} \, e^{-\frac{\beta}{2C} \mathfrak r^2} e^{\frac{\sigma}{ C } \left(\mathfrak r^2 -\mathfrak p^2\right)} \ \frac{ \Gamma\left(2\ell\pm i(\mathfrak p\pm \mathfrak r)\right) }{ \Gamma(4\ell)} ~.\label{eq.ZZZZ}
\end{align}
In writing \eqref{eq.ZZZZ} we have made use of the $\mathfrak{p}\rightarrow -\mathfrak{p}$ and $\mathfrak{r}\rightarrow -\mathfrak{r}$ symmetries of the integrand to convert the hyperbolic sine contributions into exponentials. The expression \eqref{eq.ZZZZ} is our final result for the {\it exact} one-point function.

The zero (1-dimensional) temperature limit of the answer is given by taking $\beta\to\infty$. In this limit, the $\mathfrak r$ integral condenses to $\mathfrak r=0$. Consequently,
\begin{equation}
	\langle \ZZ|\mathbb V_\ell(x,\bar x)|\ZZ\rangle_{\beta\to\infty} = \langle\mathcal O_\ell(2\sigma,0)\rangle_{\beta\to\infty} \sim \int\!\!\! d\mathfrak p^2 \, e^{2\pi\mathfrak p} \, e^{-\frac{\sigma}{ C } \mathfrak p^2} \ \frac{ \Gamma^2\left(2\ell\pm i(\mathfrak p)\right) }{ \Gamma(4\ell)} ~,\label{eq.ZZVacExp}
\end{equation}
which matches the answer earlier derived in \cite{Bagrets:2016cdf, Bagrets:2017pwq, Mertens:2017mtv, Lam:2018pvp}.

 One can instead compute the correlation functions in the {\it eigenstates}, $| E (\mathfrak l)\rangle := |\mathfrak l\rangle$ of the Schwarzian theory by first projecting out from the superposition of $|\mathfrak r\rangle$ states any state with energy less than $\mathfrak l^2/2C$. This is achieved by cutting off the $\mathfrak r$ integral as follows,
\begin{equation}\begin{aligned}\label{eq.eigenstateExpectationOrg}
	\langle\mathfrak l|\mathcal O_\ell(2\sigma,0)|\mathfrak l\rangle_{\beta\to\infty} &\sim \lim_{\beta\to\infty}\int\!\!\! d\mathfrak p^2 \!\!\!\int\limits_{|\mathfrak r|\ge |\mathfrak l|}d\mathfrak r^2 \, e^{2\pi(\mathfrak p+\mathfrak r)} \, e^{-\frac{\beta}{2C} \mathfrak r^2} e^{\frac{\sigma}{ C } \left(\mathfrak r^2 -\mathfrak p^2\right)} \ \frac{ \Gamma\left(2\ell\pm i(\mathfrak p\pm \mathfrak r)\right) }{ \Gamma(4\ell)} \\
	&\sim\mathfrak l \sinh(2\pi\mathfrak l) \int\!\!\!d\mathfrak p^2 e^{2\pi\mathfrak p-\frac\sigma{C}\left(\mathfrak p^2-\mathfrak l^2\right)} \ \frac{ \Gamma\left(2\ell\pm i(\mathfrak p\pm \mathfrak l)\right) }{ \Gamma(4\ell)}
\end{aligned}\end{equation}
One can think of this cut-off prescription as follows: an ensemble of states,
\begin{equation}
	\rho = e^{-\beta H} |\Psi\rangle \langle \Psi| \quad \underrightarrow{\scriptstyle \beta\to\infty} \quad |E_{\rm low}\rangle\langle E_{\rm low}|
\end{equation}
where $|E_{\rm low}\rangle$ is the lowest energy state that appears in the wavefunction $|\Psi\rangle$. By introducing the cut-off we work with a wavefunction whose lowest energy state is the one corresponding to $E_{\rm low} = E(\mathfrak l) = \mathfrak l^2/(2C)$. Taking $\beta\to\infty$ with the original ZZ wavefunction gives us the vacuum expectation value, \eqref{eq.ZZVacExp}, because the lowest energy state in ZZ-wavefunction corresponds to $\mathfrak l=0$.

\subsection*{Semiclassical limit}
Next, let us evaluate the integral appearing in \eqref{eq.ZZZZ} in the $C\sim a/{b^2}\to\infty$ limit using the saddle point analysis. We start by performing the following change of variables, $\mathfrak p+\mathfrak r = \frac {a\,M}{b^2}$ and $\mathfrak p-\mathfrak r = \omega$.
To evaluate this integral,  one may integrate $\omega$ exactly, noting that is precisely a  Mellin-Barnes type integral, \cite{gradshteyn2007} and then solve the remaining $M$ integral via saddle point.  Keeping only the leading terms as  $C\to\infty$ we find
\begin{align}
	M_* &= \frac{2a}{\beta}%\\%%\int \!\!\!dM \exp\left[ 2\pi\frac M{b^2} -\frac { M^2 }{ 8b^2 } \beta \mathfrak f\right] \Gamma\left(2\ell\pm i\frac M{b^2}\right) \ \Rightarrow \ 
	%\frac{M_*^2}{b^4} \int \!\!\!d\omega\, &\exp\left[-\mathfrak f \tau_c M_* \omega+\frac\beta4\mathfrak f M_* \omega\right]\, \frac{\Gamma\left(2\ell\pm i\omega\right) }{ \Gamma(4\ell)} = \frac{M_*^2}{b^4} \left( \frac1{2\sin\left( \frac{4\pi}\beta \tau_c\right)} \right)^{4\ell}
\end{align}
Finally, the semiclassical limit of the correlation function becomes
\begin{align}
	\langle \ZZ|\mathbb V_\ell(x,\bar x)|\ZZ\rangle &\sim \frac{2^{3/2}M_*^2}{2b} \, \left({8b^4}\right)^{2\ell} \frac1{2b^2} \left( 2 \pi  e^{\frac{2 C\pi^2}{\beta}} \right) \left( \frac1{2\sin\left( \frac{2\pi}\beta \sigma\right)} \right)^{4\ell}~.
\end{align}
To analyze the operator expectation value in the high energy eigenstates, let us consider \eqref{eq.eigenstateExpectationOrg} with $\mathfrak l\sim\mathcal O (C)$. In this case, consider $\mathfrak p = \mathfrak l+m$,
\begin{align}
	\langle\mathfrak l|\mathcal O_\ell(2\sigma,0)|\mathfrak l\rangle_{\substack{\beta\to\infty\\\mathfrak l \sim C}} &\sim \mathfrak l \sinh(2\pi\mathfrak l) \!\!\!\int\!\!\!d m\, (\mathfrak l +m) e^{2\pi(\mathfrak l+m)-\frac\sigma{C}(2\mathfrak l m+m^2)} \ \frac{ \Gamma\left(2\ell\pm i(2 \mathfrak l+m)\right)  \Gamma\left(2\ell\pm im\right) }{ \Gamma(4\ell)} \nonumber\\
	&\sim \mathfrak l^2 \sinh(2\pi\mathfrak l) {\left(2\mathfrak{l}\right)^{4 \ell -1}} \int\!\!\!d m\, e^{\pi m-\frac{2\sigma\mathfrak l m}{C}} \ \frac{ \Gamma\left(2\ell\pm im\right) }{ \Gamma(4\ell)} \nonumber\\
	&\sim \left(\frac12 \frac1{i \sin\left( \frac{2\pi\mathfrak l}C \sigma \right)}\right)^{4\ell} \label{eq.eigenstateExpectation}
\end{align}
The effective temperature due to the heavy state is given by, $T_{\textrm{eff}} = { \mathfrak l}/C = \sqrt{2E(\mathfrak l)/C}$. This is consistent with the answer that we obtained in \eqref{eq.SchwarzSaddleExp}.
\subsubsection{Two-point function of light operators \texorpdfstring{$\langle \ZZ| \mathbb V_{\ell_2}(x_2,\bar x_2) \mathbb V_{\ell_1}(x_1,\bar x_1) |\ZZ\rangle$}{<ZZ|V2 V1|ZZ>}}
We have now shown that the insertion of ZZ branes in the 2D picture allows us to study Schwarzian expectation values either at finite temperature, or using the projection trick, with respect to the eigenstates of the theory. We now move on to higher-point correlators of operators of weight ${\cal O}(1)$ and the interesting physics associated to them. To this end, we study the time ordered correlation function, $\langle \ZZ| \mathbb V_{\ell_2}(x_2,\bar x_2) \mathbb V_{\ell_1}(x_1,\bar x_1) |\ZZ\rangle$. Because of the dimensional reduction combined with the doubling trick described in section \ref{sec.stateInterpretation}, from the 1-dimensional point of view we get a configuration of bilocal operator insertion with $-\pi<-\sigma_2<-\sigma_1<0<\sigma_1<\sigma_2<\pi$. In the Liouville theory the expression of this correlation function is,
\begin{align}
	\langle \ZZ|\mathbb V_{\ell_2}(x_2,\bar x_2)  \mathbb V_{\ell_1}(x_1,\bar x_1) |\ZZ\rangle&=2^{3/2} \,\pi^2\!\!\!\int\!\!\! dP^2 \,dR^2 \, \frac{ \left(\pi\mu\gamma(b^2)\right)^{i\frac{P-R}b} }{ \Gamma\left(1+2ibP\right) \Gamma\left(1+2i\frac Pb\right) } \nonumber\\
	&\hspace{3cm} \times \frac{\left\langle\nu_R\right\vert \mathbb V_{{\ell}_2}(x_2,\bar x_2) \mathbb V_{\ell_1}(x_1,\bar x_1) \left\vert \nu_P\right\rangle }{ \Gamma\left(1-2ibR\right) \Gamma\left(1-2i\frac Rb\right) }
\end{align}
which, in 1-dimension, reduces to,
\begin{align}\label{eq.4-pt-exact}
	&\sim\!\!\int\!\!\! d\mathfrak p^2 \,d\mathfrak r^2 d\mathfrak d^2 \, \sinh\left(2\pi \mathfrak d\right)\sinh \left( 2\pi \mathfrak p \right) \sinh \left( 2\pi \mathfrak r \right) \frac{\Gamma\left( 2{\ell}_2 \pm i (\mathfrak r \pm \mathfrak d) \right)}{\Gamma(4{\ell}_2)}\, \frac{\Gamma\left( 2{\ell}_1 \pm i (\mathfrak d \pm \mathfrak p) \right)}{\Gamma(4{\ell}_1)}  \nonumber \\% 2^{3/2-2} \, \frac{\pi^2}{\pi^2} \left(-\frac{1}{b\pi}\right)
	&\hspace{3cm}\times\exp \left[ -\frac{1}{2 C} \left( \beta \mathfrak{r}^2  +2 \sigma_2(\mathfrak d^2-\mathfrak r^2) +2\sigma_1(\mathfrak p^2-\mathfrak d^2)\right) \right]
\end{align}

The Lyapunov exponent for the Schwarzian theory in the thermal ensemble has already been computed from the OTO two-point function of bilocals in \cite{Mertens:2017mtv,Lam:2018pvp} using the ZZ-brane perspective. Here our main interest is in using the projection trick of section \ref{sec.BilocalOnePointZZ} to instead  compute the Lyapunov exponent with respect to the eigenstate density operator $|E\rangle \langle E |$, which is the main object of interest in this paper. Applying the projection trick discussed previously to \eqref{eq.4-pt-exact} results in the expression,
\begin{align}
	&\sim \lim_{\beta\to\infty}\int\!\!\! d\mathfrak p^2 d\mathfrak d^2 \!\!\int\limits_{|\mathfrak r|\ge |\mathfrak l|} \,d\mathfrak r^2  \, \sinh\left(2\pi \mathfrak d\right)\sinh \left( 2\pi \mathfrak p \right) \sinh \left( 2\pi \mathfrak r \right) \frac{\Gamma\left( 2{\ell}_2 \pm i (\mathfrak r \pm \mathfrak d) \right)}{\Gamma(4{\ell}_2)}\, \frac{\Gamma\left( 2{\ell}_1 \pm i (\mathfrak d \pm \mathfrak p) \right)}{\Gamma(4{\ell}_1)}  \nonumber \\% 2^{3/2-2} \, \frac{\pi^2}{\pi^2} \left(-\frac{1}{b\pi}\right)
	&\hspace{3cm}\times\exp \left[ -\frac{1}{2 C} \left( \beta \mathfrak{r}^2  +2 \sigma_2(\mathfrak d^2-\mathfrak r^2) +2\sigma_1(\mathfrak p^2-\mathfrak d^2)\right) \right]\nonumber\\
	&\sim \, \!\! \mathfrak{l} \sinh \left( 2\pi \mathfrak l \right) \int\!\!\! d\mathfrak p^2  d\mathfrak d^2 \, \sinh\left(2\pi \mathfrak d\right)\sinh \left( 2\pi \mathfrak p \right)  \frac{\Gamma\left( 2{\ell}_2 \pm i (\mathfrak l \pm \mathfrak d) \right)}{\Gamma(4{\ell}_2)}\, \frac{\Gamma\left( 2{\ell}_1 \pm i (\mathfrak d \pm \mathfrak p) \right)}{\Gamma(4{\ell}_1)}  \nonumber \\% 2^{3/2-2} \, \frac{\pi^2}{\pi^2} \left(-\frac{1}{b\pi}\right)
	&\hspace{3cm}\times\exp \left[ -\frac{1}{2 C} \left( 2 \sigma_2(\mathfrak d^2-\mathfrak l^2) +2\sigma_1(\mathfrak p^2-\mathfrak d^2)\right) \right]
\end{align}
In order to make contact with our semi-classical results on chaos in eigenstate \ref{sec.SemiclassicalChaos} we also need an expression for the OTO version of the above. This can be formally achieved by insertion the $R$-matrix of \cite{Ponsot:1999uf}, analogous to the thermal case worked out in \cite{Mertens:2017mtv},
\begin{align}\label{eq.R-mat-OTOC-exact}
	\langle\ZZ|& \mathbb V_{\ell_2}(x_2,\bar x_2) \mathbb V_{\ell_1}(x_1,\bar x_1) |\ZZ\rangle_{\rm OTOC} = \nonumber\\
	& \qquad \int \!\!\!d \mathfrak p d\mathfrak r d \mathfrak p_s d \mathfrak p_t \, C(-\mathfrak r ,2\ell_2, \mathfrak p_s) C(-\mathfrak p_s, 2\ell_1,\mathfrak p) \, \Psi_\ZZ (b\mathfrak r) \Psi_\ZZ(-b\mathfrak p) \nonumber \\
	&\hspace{3cm} \times R_{{}_{\mathfrak p_s\, \mathfrak p_t}} \left[\, {}^{2\ell_2}_{\mathfrak r} \,{}^{2\ell_1}_{\mathfrak p}\right]  \mathcal{F}_{\mathfrak p_t} \left[\, {}^{2\ell_1}_{\mathfrak r} \,{}^{2\ell_2}_{\mathfrak p} ; \frac{x_1}{x_2} \right] \mathcal{F}_{\mathfrak p_s} \left[\, {}^{2\ell_2}_{\mathfrak r} \,{}^{2\ell_1}_{\mathfrak p} ; \frac{\bar x_2}{\bar x_1} \right]%e^{- \frac{x_2}{x_1} \mathfrak p_{t}^2} \, 
\end{align}
where $C(y_1, y_2, y_3)$ is the DOZZ three-point function, \cite{Dorn:1994xn, Zamolodchikov:1995aa}.
After reduction to 1-dimension we obtain a final answer,
\begin{equation}\begin{aligned}
	\frac{\langle O_{\ell_2}(\sigma_3,\sigma_4) O_{\ell_1}(\sigma_1,\sigma_2)\rangle}{\langle O_{\ell_2}(\sigma_3,\sigma_4)\rangle\langle O_{\ell_1}(\sigma_1,\sigma_2)\rangle}\sim \int \!\!\!d \mathfrak p^2 d\mathfrak r^2 d \mathfrak p_s^2 d \mathfrak p_t^2 \, \sinh(2\pi\mathfrak p) \sinh(2\pi\mathfrak r) \sinh(2\pi\mathfrak p_s) \sinh(2\pi\mathfrak p_t)\\
	\times R_{\mathfrak p_s \mathfrak p_t}\left[\, {}^{2\ell_2}_{\mathfrak r} \,{}^{2\ell_1}_{\mathfrak p}\right] \exp\left[-\frac1{2C}\left({\mathfrak p^2}(\beta-\sigma_{41})-\mathfrak p_t^2 \sigma_{31}-\mathfrak r^2 \sigma_{32}-\mathfrak p_s^2 \sigma_{41}\right)\right]\\
	\times \frac{\sqrt{\Gamma(2\ell_1\pm i(\mathfrak p\pm \mathfrak p_t)) \Gamma(2\ell_2\pm i(\mathfrak r\pm \mathfrak p_t)) \Gamma(2\ell_1\pm i(\mathfrak r\pm \mathfrak p_s)) \Gamma(2\ell_2\pm i(\mathfrak p\pm \mathfrak p_s))}}{\Gamma(4\ell_1)\Gamma(4\ell_2)}
\end{aligned}\end{equation}
In the above and subsequent equations, we have generalized the placement of the operators to arbitrary points $\sigma_i$, with the ordering, $-\beta/2<\sigma_1<\sigma_2<\sigma_3<\sigma_4<\beta$. The R-matrix, $R_{\mathfrak p_s \mathfrak p_t}\left[\, {}^{2\ell_2}_{\mathfrak r} \,{}^{2\ell_1}_{\mathfrak p}\right]$ is related to the 6-j symbols of $SL(2,\mathbb R)$. Its expression is rather daunting and we refer the reader to Appendix B of \cite{Mertens:2017mtv}.

However, we can bypass this procedure by appealing to our results in section \ref{sec.AlternativeDerviation} and express the OTO correlation function in terms of the identity Virasoro block

\be
\frac{\langle {\rm O}_{\ell_1}(\sigma_1, \sigma_2){\rm O}_{\ell_2}(\sigma_3, \sigma_4) \rangle}{\langle {\rm O}_{\ell_1}(\sigma_1, \sigma_2) \rangle\langle  {\rm O}_{\ell_2}(\sigma_3, \sigma_4)\rangle} \sim y^{2 \Delta_{\ell_1}}  \mathcal{F}_{\rm id} \left[\, {}^{\Delta_{\ell_1}}_{\Delta_{\ell_1}} \,{}^{\Delta_{\ell_2}}_{\Delta_{\ell_2}} ; y \right] \,,
\ee
where $y = 1-z$, as defined in \eqref{eq.identityBlock} and \eqref{eq.CrossRatio} above.
Note that we had to pass to the semiclassical limit $C\rightarrow \infty$ in order to establish this result. We see, once more, that the large-$C$ expansion of the full result \eqref{eq.R-mat-OTOC-exact} agrees with our direct semiclassical evaluation, confirming that the eigenstates of the Schwarzian are maximally scrambling with Lyapunov exponent $2\pi / \beta_{\rm eff}$.

\subsubsection{Heavy-light two-point function \texorpdfstring{$\langle \ZZ|\mathbb V_{\ell^H}(x_1,\bar x_1) \mathbb V_{\ell^L}(x_2,\bar x_2)|\ZZ\rangle$}{<ZZ|V-HV-L|ZZ>}}
We next want to study the effect of inserting heavy operators on the (effective) temperature as perceived by light operators. We evaluate the 4-point function, \eqref{eq.4-pt-exact}, with a heavy operator, $\mathbb V_{\ell_1} = \mathbb V_{\ell^H}$ with $\ell^H\sim C$ and a light operator $\mathbb V_{\ell_2} = \mathbb V_{\ell^L}$ with $\ell^L\sim 1$, in the classical limit to study ETH. Since the dimensionless parameter in our problem is $C/\beta$, the classical zero temperature limit can be taken in two ways: $C/\beta\to\infty$ with $\beta$ fixed, followed by $\beta\to\infty$; or, by taking $C,\beta\to\infty$ simultaneously such that $\mathfrak C= C/\beta$ is finite followed by $\mathfrak C\to\infty$. We find that the thermal behaviour of this 4-point function depends on the order of limits.
\subsection*{$C/\beta\to\infty$ with $\beta$ fixed, followed by $\beta\to\infty$}
\begin{figure}[t]
    \centering
    \subfloat[$\omega$]{{\includegraphics[trim={0 0 4.9cm 0},clip,width=0.40 \textwidth]{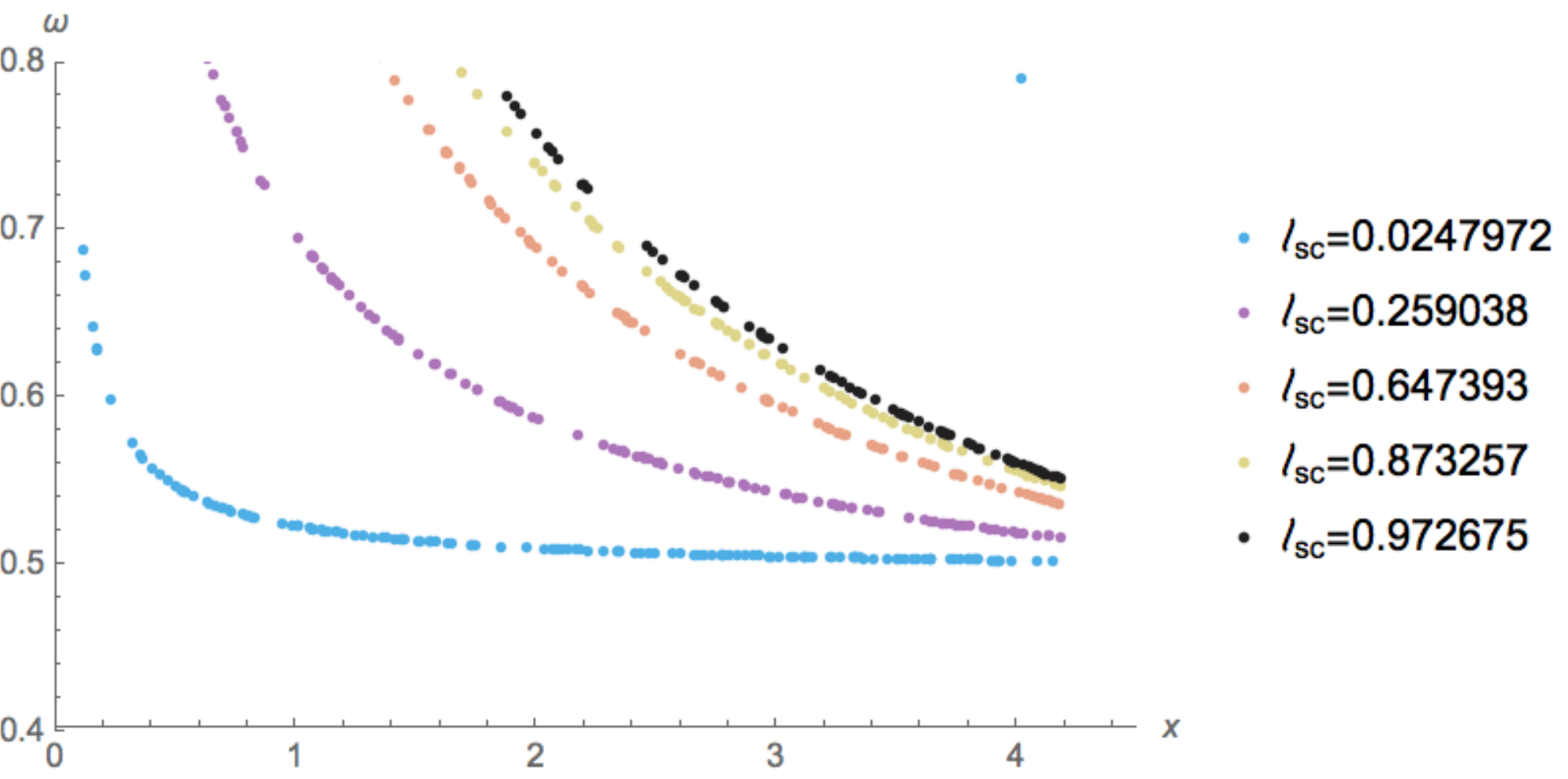} }}%
    \qquad
    \subfloat[$\tilde\omega$]{{\includegraphics[width=0.52 \textwidth]{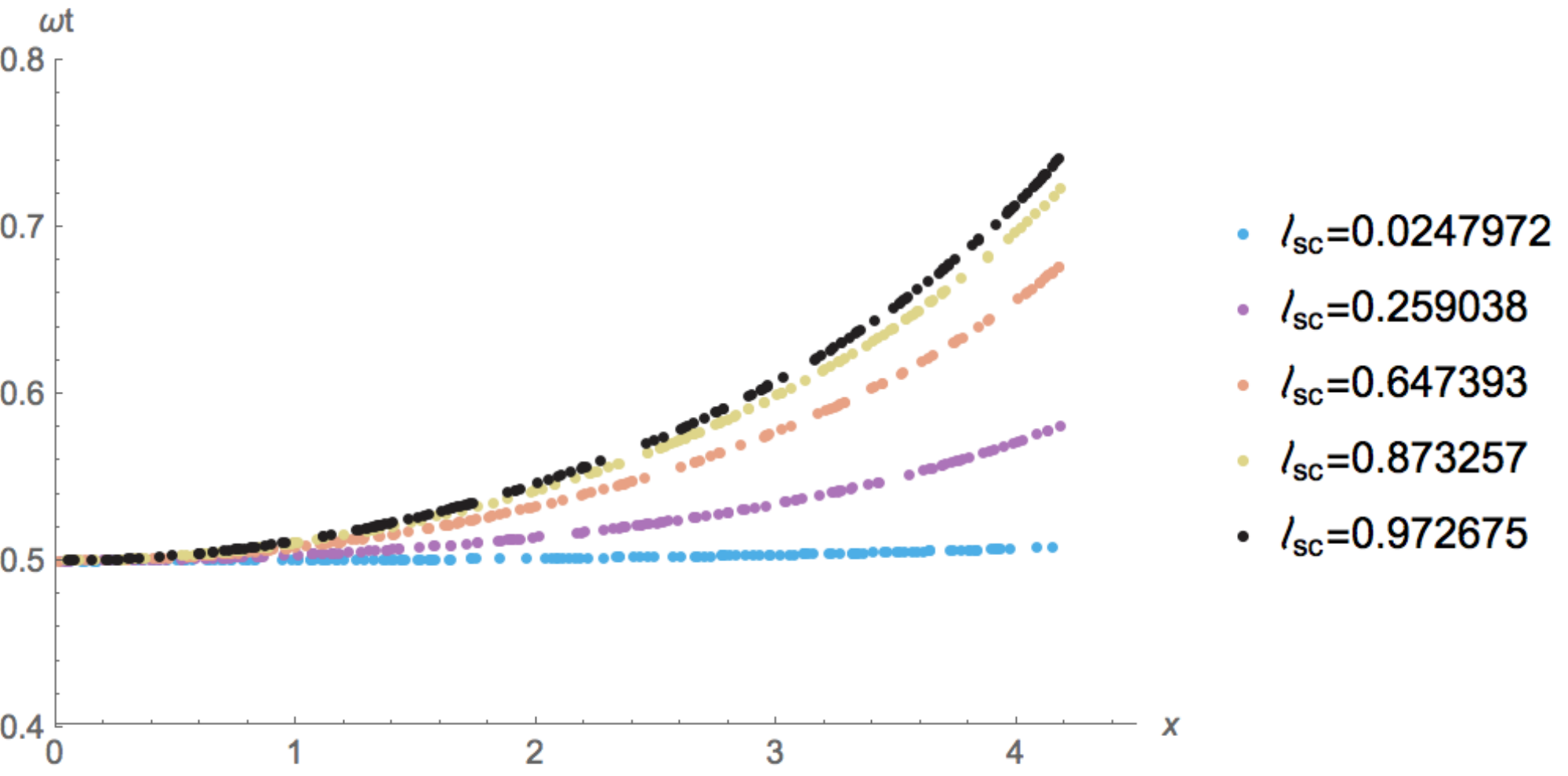} }}%
    \caption{Plots for increasing values of $\ell_{sc}$ demonstrate that the solution for $\omega,\tilde\omega$ for a given $x$ increases with increasing $\ell_{sc}$. In the above plot, $C = 567.605, ~ \beta=2\pi$. The solution for $x=\beta$ is $\omega=\pi/\beta$ for all values of $\ell_{sc}$. While the solution for $x=0$ is $\tilde\omega=\pi/\beta$ for all values of $\ell_{sc}$.}%
    \label{fig:om-omt-v-x}%
\end{figure}

In this limit, we perform a change of variables $\mathfrak r^2 = \mathfrak d^2 +m$. Consequently, the $\mathfrak p,\mathfrak d$ integrals are performed using saddle point analysis while the $m$ integral is done using the Mellin-Barnes technique discussed above. 
\begin{align}
	&\sim \int\!\!  d\mathfrak{p}^2 \, d \mathfrak{d}^2\, \mathfrak{d}^{ 4\ell^L} \sinh \left(2 \pi  \mathfrak{p} \right) \sinh \left( 2\pi \mathfrak{d} \right) \frac{\Gamma\left( 2{\ell}^H \pm i (\mathfrak{p} \pm \mathfrak{d}) \right)}{\Gamma(4{\ell}^H)} \nonumber \\
		&\qquad \quad  \quad \quad \times    \exp \left[ -\frac{1}{2 C} \left( \beta \mathfrak{d}^2 + 2\sigma_1 \left( \mathfrak p^2-\mathfrak d^2\right)\right) \right] \times  \left[\frac{\mathfrak{d}}{\sin\left( \frac{\mathfrak{d}}{2C}\left(\beta-2\sigma_2\right) \right)} \right] ^{4\ell^L}
\end{align}
The saddle point equations, in the variables $\mathfrak p$ and $\mathfrak d$, of the above integral are,
\begin{subequations}\label{eq:FinalMonEqs}\begin{align}
	\label{eq:FinalMonEq1} \ \ \tan(x\omega) &= -\frac{4 \omega \ell_{sc} }{4\tilde\omega^2-4 \omega ^2+\ell_{sc}^2} \qquad \text{assuming $ \qquad 4 \tilde\omega ^2 - 4 \omega^2 + \ell_{sc}^2 \neq0$} \\
	\label{eq:FinalMonEq2}  \ \ \tan \Big(\tilde\omega (\beta -x ) \Big) &= -\frac{4 \tilde\omega \ell_{sc} }{4 \omega ^2-4 \tilde\omega^2+\ell_{sc}^2} \qquad \text{assuming $\qquad 4 \omega^2 -4 \tilde\omega ^2 + \ell_{sc}^2 \neq0$}~.
\end{align}\end{subequations}
For simplicity, here we have scaled the variables as follows,
\begin{equation}
	\omega = \frac{\mathfrak d}{ 2C }, \quad \ \tilde \omega = \frac{\mathfrak p}{ 2C }, \quad \ \ell_{sc} = \frac{2\ell^H}C, \quad \ x = 2\sigma_1~.
\end{equation}
While these equations are transcendental and thus not exactly solvable, we can find the solutions for $\omega, \tilde \omega$ is special limits. Note that the solutions of these variables depend on the separation of the heavy insertions, $x$. We take $x=\beta/2$. At this value of $x$, both equations are equivalent to
\begin{equation}\label{eq:selfDualPoint}
	\tan\left(\frac\beta2\omega\right) = -\frac{4 \omega }{\ell_{sc}}~.
\end{equation}
For very large values of $\ell_{sc}$, the solutions of this equation are given by,
\begin{align}\label{eq:Teff}
	\omega_n &= 2 n \frac{\pi}{\beta} - \mathfrak c_n~,\quad \text{where,	} \mathfrak c_n = \frac{16 \pi  n}{\beta  (\beta  \ell_{sc}+8)}~,\quad n\in\mathbb Z~,\nonumber\\[10pt]
	\Rightarrow T_{\rm eff} &= \frac1{\beta_{\rm eff}} = \frac{\omega_n}{\pi} =  \frac{2 n T \ell_{sc}}{ \ell_{sc} + 8 T}~.
\end{align}
Here, $T=1/\beta$ is the background temperature that we began with; $T_{\rm eff} = 1/\beta_{\rm eff}$ is the effective temperature perceived by the light operator in the presence of the heavy insertions,
\begin{equation}\label{eq:normalized4-pt}\begin{aligned}
		\frac{ \langle\mathcal O_{\ell^H} \mathcal O_{\ell^L} \mathcal O_{\ell^L} \mathcal O_{\ell^H}\rangle}{\langle\mathcal O_{\ell^H}\mathcal O_{\ell^H} \rangle} \sim \left(\frac\pi{ \beta_{\rm eff}~\sin\left(\frac {2\pi\sigma_2}{\beta_{\rm eff}} \right) }\right)^{4\ell^L}~.
\end{aligned}\end{equation}%
The $\ell_{sc}\to0$ limit leaves only $n=1$ as a physical solution; for all other solutions, $\ell_{sc}\to0$ limit leaves an unphysical `residual' temperature. An observant reader may point out that for low enough values of $\ell_{sc}$, $T_{\rm eff}<T$, but that is outside the validity of the above approximation. Moreover, the asymptotic temperature at large $\ell_{sc}$ provides an upper-bound on the effective temperature. This is simply because we do not expect an operator with smaller $\ell_{sc}$ to create a thermal state at a higher temperature than an operator with higher $\ell_{sc}$. The more general solution for the effective temperature needs to be computed numerically, and can be inferred from \autoref{fig:om-omt-v-x}.
Note that the effective temperature due to the presence of the heavy operators is always greater than the background temperature, $T=1/\beta$ that we start with. However, the effective temperature uniformly approaches zero as the background temperature $T\to0$. This is \emph{similar} to the result for the FZZT brane that we obtained in \autoref{sec.Semiclassics} in that the effective temperature is proportional to the background temperature.

\subsection*{$\mathfrak C= C/\beta$ is finite followed by $\mathfrak C\to\infty$}
Correlation functions in this other limit were studied in \cite{Lam:2018pvp} and we include their result for completeness of presentation. In their analysis they found that the four point function (two point function of the bilinears) does not show ETH, as the correlation function actually becomes periodic in real time. Written in Euclidean time, they find \cite{Lam:2018pvp}
\begin{equation}\begin{aligned}
		\frac{ \langle\mathcal O_{\ell^H} \mathcal O_{\ell^L} \mathcal O_{\ell^L} \mathcal O_{\ell^H}\rangle }{\langle \mathcal O_{\ell^H}\mathcal O_{\ell^H} \rangle} \sim \left(\frac\pi{ \beta_{\rm eff}~\sinh\left(\frac {2\pi\sigma_2}{\beta_{\rm eff}} \right) }\right)^{2\ell^L}, \qquad \beta_{\rm eff} = \frac { 2\pi }{\ell_{sc}}~.
\end{aligned}\end{equation}
Drawing analogy with AdS$_3$/CFT$_2$ correspondence, this correlation function behaves like a correlation function computed in thermal AdS$_3$.  It would be interesting to obtain an analogous bulk interpretation for the case of AdS$_2$.

\subsection{Computations between ZZ and FZZT branes}
We now use the relationship between 2D Liouville theory and the Schwarzian in the presence of FZZT branes. This allows us to write down the full Schwarzian correlation functions in the presence of coherent states of the type \eqref{eq.SchwarzCoherent}. As we shall see, their leading semiclassical limits will once more confirm our previous results obtained by other means.

\subsubsection{Bilocal one-point function \texorpdfstring{$\langle \FZZT|\mathbb V_\ell(x,\bar x)|\ZZ\rangle$}{<\FZZT|V-l|\ZZ>}}
In this section we compute the special case of \eqref{eq.ZZFZZSchToState} with a single operator insertion using the FZZT-wavefunction given by \eqref{eq.ZZFZZTwavefunctions},
\be
	\Psi_{FZZT}^{(s)}(P) := \langle \FZZT|P\rangle = e^{2i\pi P s} \left[-\frac{i}{2\pi P} \left(\frac{\gamma(b^2)}{8b^2}\right)^{-\frac{iP}b} \Gamma\left(1+2iPb\right)\Gamma\left(1+2\frac{iP}b\right)\right]
\ee
Recall, the variable $s$ is related to the parameter $r$ that labels the FZZT branes by equation \eqref{eq.relationBWparas}. We are now ready to put together all the ingredients to compute the expectation value of a bilocal operator with respect to the FZZT boundary state. We start by writing down the full expression for $\langle \mathbb V_\ell(x,\bar x) \rangle_r := \langle \FZZT|\mathbb V_\ell(x,\bar x)|\ZZ\rangle$, which takes the form
\begin{align}
	\langle \mathbb V_\ell(x,\bar x) \rangle_r = -2^{3/4}\int\!\! dP \,dR \,\frac PR e^{2\pi i R s} \frac{ \left(\frac{\gamma(b^2)}{8b^2}\right)^{i\frac{P-R}b} \Gamma\left(1+2ibR\right) \Gamma\left(1+2i\frac Rb\right) }{ \Gamma\left(1+2ibP\right) \Gamma\left(1+2i\frac Pb\right) } \left\langle\nu_R\right\vert \mathbb V_\ell(x,\bar x)\left\vert \nu_P\right\rangle
\end{align}
This expression differs from the pure ZZ one only in the measure factor where, roughly speaking, a ZZ brane corresponds to an insertion of $\sinh(2\pi \mathfrak p)$, while an FZZT brane corresponds to an insertion of $\cos(2\pi \mathfrak p)$.
Once again, using \eqref{eq.dozz}, and performing the same kind of manipulations as for the pure ZZ case above, we can write,
\begin{align}
	\langle \mathbb V_\ell(x,\bar x) \rangle_r  &= -\frac{2^{3/4}}{2 b} \left({8b^4}\right)^{2\ell} \frac{b^2}\pi e^{-\frac{\pi \beta Q^2}{ 4a }} \int\!\! d\mathfrak p^2 \,d\mathfrak r \, e^{2\pi \left(i \mathfrak r \hat s+\mathfrak p\right)} \, e^{-\frac{\beta}{2C} \mathfrak r^2} e^{\frac{\sigma}{ C } \left(\mathfrak r^2 -\mathfrak p^2\right)} \ \frac{ \Gamma\left(2\ell\pm i(\mathfrak p\pm \mathfrak r)\right) }{ \Gamma(4\ell)}~.\label{eq.FZZTZZ2}
\end{align}
Above, we have introduced a rescaled $\hat s = s\, b$ parameter for the FZZT brane. This scales the energy of the FZZT brane in an appropriate fashion from the 1D perspective. This can be easily seen by recalling that $s$ is related to the parameter $r$ in the boundary term, \eqref{eq.NeumannBdy} by,
\begin{equation}\begin{aligned}
	\cosh^2(\pi b s) &= \frac{r^2}{2\pi b^2} \sin(\pi b^2)\\
	&\Big\downarrow b\to0\\
	1+&\frac{\pi^2b^2s^2}2 \approx \frac{r^2}2
\end{aligned}\end{equation}
In the $b\to0$ limit, RHS of the above equation is an $\mathcal O(1)$ parameter, therefore $s\sim\frac 1b$ which justifies our redefinition, $s=\hat s/b$.

Evaluating \eqref{eq.FZZTZZ2} using the saddle point integration method described above we get,
\begin{equation}
	\langle \FZZT|\mathbb V_\ell(x,\bar x)|\ZZ\rangle \sim \left( \frac1{2\sinh\left( \frac{2\pi}\beta \hat s \sigma \right)} \right)^{4\ell}~.
\end{equation}
Since $\sigma$ is indeed Euclidean time in the Schwarzian theory, this result is thermal for the orbit when $\hat s \in i \mathbb{R}$ and non-thermal if $\hat s \in \mathbb{R}$. In the former case it is thermal at the effective temperature $\beta_{\rm eff} = \beta / \hat{s}$. A similar result in 2D CFT was recently found in \cite{David:2019bmi,Jensen:2019cmr}. 

We would also like to compute exact correlation function akin to \autoref{eq.ZZFZZTsemicalssical} above using these methods from 2D Liouville theory. However, the asymmetry of the configuration does not render it immediately amenable to employing the doubling trick and requires the development of the Feynman rules\footnote{Such Feynman rules are derived for the pure ZZ-ZZ case in \cite{Mertens:2017mtv}.} in the one-dimensional theory in the presence of insertion of the FZZT operators discussed in section \ref{sec.descent}. We will investigate these questions in our future work.

\subsubsection{Bilocal two-point function \texorpdfstring{$\langle \FZZT| \mathbb V_{\ell_2}(x_2,\bar x_2) \mathbb V_{\ell_1}(x_1,\bar x_1)|\ZZ\rangle$}{<FZZT|[V1,V2]|ZZ>}}
Like in the previous case, we next compute the 2-point correlation function of the bilocal operators. From the perspective of 2-dimensional Liouville theory it is given by,
\begin{align}
	\langle \FZZT|\mathbb V_{\ell_2}(x_2,\bar x_2) &\mathbb V_{\ell_1}(x_1,\bar x_1) |\ZZ\rangle =-2^{3/4} \, \!\!\!\int\!\!\! dP \,dR \frac{P}{R} e^{2\pi i R s} \left(\pi\mu\gamma(b^2)\right)^{i\frac{P-R}b}\nonumber \\
	&\ \times \frac{ \Gamma\left( 1 + 2ibR \right)\Gamma\left( 1 + 2i \frac{R}{b} \right) }{ \Gamma\left(1+2ibP\right) \Gamma\left(1+2i\frac Pb\right) } \left\langle\nu_R\right\vert \mathbb V_{{\ell}_2}(x_2,\bar x_2) \mathbb V_{\ell_1}(x_1,\bar x_1) \left\vert \nu_P\right\rangle ~,
\end{align}
and in 1-dimension, this expression reduces to,
\begin{align}
	&\sim\!\!\int\!\!\! d\mathfrak r d\mathfrak p^2 \,d\mathfrak d^2   \, e^{2\pi i \mathfrak r \hat{s}} \sinh\left(2\pi \mathfrak p\right)\sinh \left( 2\pi \mathfrak d \right) \frac{\Gamma\left( 2{\ell}^2 \pm i (\mathfrak r \pm \mathfrak d) \right)}{\Gamma(4{\ell}^2)}\, \frac{\Gamma\left( 2{\ell}^1 \pm i (\mathfrak d \pm \mathfrak p) \right)}{\Gamma(4{\ell}^1)}  \nonumber \\% 2^{3/2-2} \, \frac{\pi^2}{\pi^2} \left(-\frac{1}{b\pi}\right)
	&\hspace{3cm}\times\exp \left[ -\frac{1}{2 C} \left( \beta \mathfrak{r}^2  +2 \sigma_2(\mathfrak d^2-\mathfrak r^2) +2\sigma_1(\mathfrak p^2-\mathfrak d^2)\right) \right]~.
\end{align}
Moreover, we can also compute the out-of-time-ordered correlation functions using the R-matrices. The exact expression for such an OTOC is given by,
\begin{equation}\begin{aligned}\label{eq.R-mat-OTOC-exact-ZZFZZ}
	\frac{\langle O_{\ell_2}(\sigma_3,\sigma_4) O_{\ell_1}(\sigma_1,\sigma_2)\rangle}{\langle O_{\ell_2}(\sigma_3,\sigma_4)\rangle\langle O_{\ell_1}(\sigma_1,\sigma_2)\rangle}\sim \int \!\!\!d \mathfrak p^2 d\mathfrak r d \mathfrak p_s^2 d \mathfrak p_t^2 \, \sinh(2\pi\mathfrak p) \cos(2\pi\mathfrak r \hat s) \sinh(2\pi\mathfrak p_s) \sinh(2\pi\mathfrak p_t)\\
	\times R_{\mathfrak p_s \mathfrak p_t}\left[\, {}^{2\ell_2}_{\mathfrak r} \,{}^{2\ell_1}_{\mathfrak p}\right] \exp\left[-\frac1{2C}\left({\mathfrak p^2}(\beta-\sigma_{41})-\mathfrak p_t^2 \sigma_{31}-\mathfrak r^2 \sigma_{32}-\mathfrak p_s^2 \sigma_{41}\right)\right]\\
	\times \frac{\sqrt{\Gamma(2\ell_1\pm i(\mathfrak p\pm \mathfrak p_t)) \Gamma(2\ell_2\pm i(\mathfrak r\pm \mathfrak p_t)) \Gamma(2\ell_1\pm i(\mathfrak r\pm \mathfrak p_s)) \Gamma(2\ell_2\pm i(\mathfrak p\pm \mathfrak p_s))}}{\Gamma(4\ell_1)\Gamma(4\ell_2)}
\end{aligned}\end{equation}

\noindent Once again, we bypass the exact computation of this integral and resort to the analysis of section \ref{sec.SemiclassicalChaos} to argue that the Lyapunov exponent in this scenario will be given by $\lambda = 2\pi/\beta_{\rm eff}$ where $\beta_{\rm eff} = \beta/\hat s$.

\subsection{Computations between two FZZT branes}
Before moving on to the discussions section, for completeness we would like to gather exact results for correlation functions in the presence of FZZT branes at both ends, proceeding along similar lines as above.
\subsubsection{\texorpdfstring{$\langle \FZZT|\mathbb V_\ell(x,\bar x)|\FZZT\rangle$}{<FZZT|V-l|FZZT>}}
\begin{figure}
\begin{center}
	\includegraphics[width=0.6\textwidth]{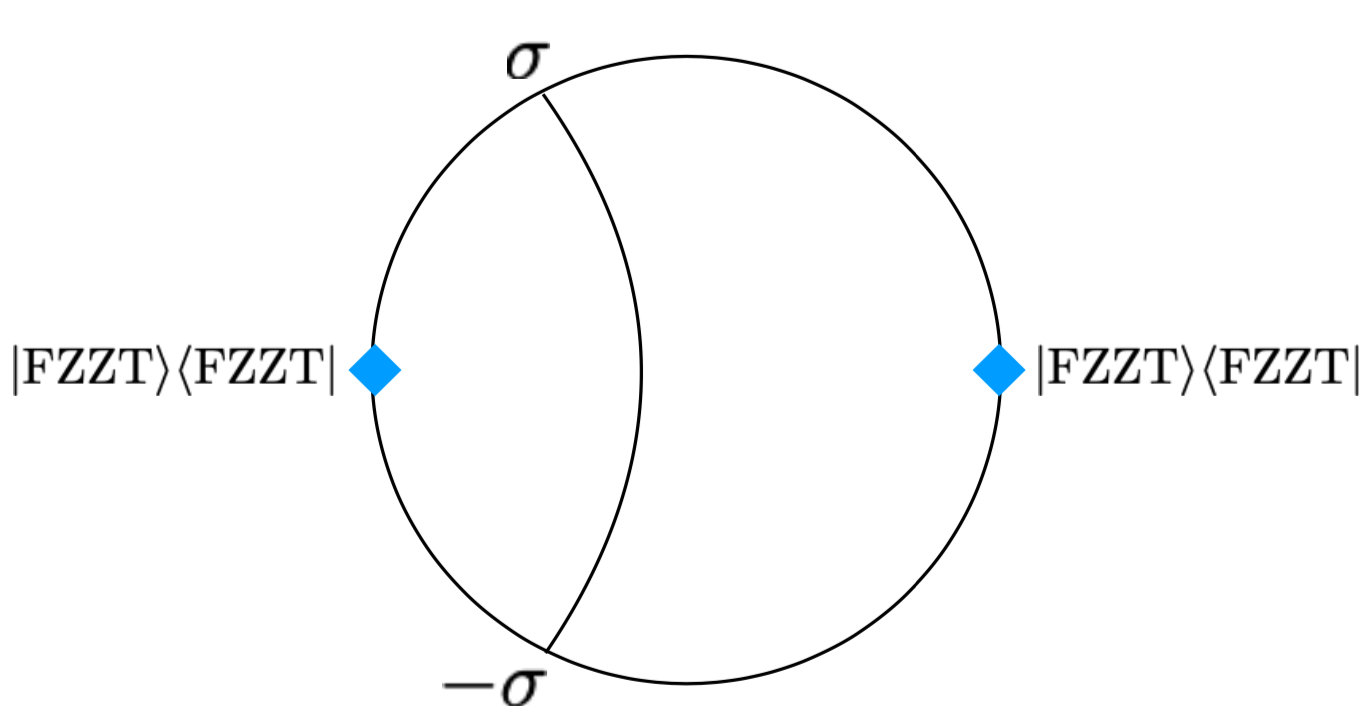}
\caption{The configuration of the insertions in the case of two FZZT operators.}
\label{fig.2-fzzt}
\end{center}
\end{figure}

This correlation function corresponds to one with two insertions of FZZT operators as shown in \autoref{fig.2-fzzt}.
Using the FZZT wavefunctions in \eqref{eq.ZZFZZTwavefunctions}, we can write this one-point function of the bilocal operator as follows,
\begin{align}
	\langle \FZZT|\mathbb V_\ell(x,\bar x)|&\FZZT\rangle=\frac1{4\pi^2}\int\!\! dP \,dR \,\frac 1{PR} e^{2\pi i (s\,R-s'\,P)} \left(\frac{\gamma(b^2)}{8b^2}\right)^{i\frac{P-R}b} \nonumber \\
	&\hspace{-0.5cm}\times \left[\Gamma\left(1+2ibR\right) \Gamma\left(1+2i\frac Rb\right) \Gamma\left(1-2ibP\right) \Gamma\left(1-2i\frac Pb\right) \right] \left\langle\nu_R\right\vert \mathbb V_\ell(z,\bar z)\left\vert \nu_P\right\rangle~.
\end{align}
Once again, using \eqref{eq.dozz}, we can write,
\begin{align}%
%	\langle \FZZT|\mathbb V_\ell(x,\bar x)|\FZZT\rangle &= \frac1{2b} \left({8b^4}\right)^{2\ell} \frac1{\pi^2}\int\!\! d\mathfrak p \,d\mathfrak r \left\{ \left(b^2\gamma(b^2)\right)^{i(\mathfrak p-\mathfrak r)} \Gamma\left(1+2ib^2\mathfrak r\right) \Gamma\left(1-2ib^2 \mathfrak p\right) \right\} \nonumber \\
%	&\times e^{2\pi i (\hat s\,\mathfrak r-\hat s'\,\mathfrak p)} \frac{ \Gamma\left(2\ell\pm i(\mathfrak p\pm\mathfrak r)\right) }{ \Gamma(4\ell)}\, e^{-\frac{2\beta\pi}a \left(\frac{Q^2}4+b^2\mathfrak r^2\right)} e^{\frac{4\pi}a \sigma b^2 \left(\mathfrak r^2 -\mathfrak p^2\right)}           \\
	\langle \FZZT|\mathbb V_\ell(x,\bar x)|\FZZT\rangle&= \frac1{2b} \left({8b^4}\right)^{2\ell} \frac1{\pi^2}e^{-\frac{2\beta\pi}a \frac{Q^2}4}\int\!\! d\mathfrak p \,d\mathfrak r \, \cos(2\pi\hat s \mathfrak r) \cos(2\pi\hat s' \mathfrak p) \nonumber \\
	&\hspace{2.5cm}\times \frac{ \Gamma\left(2\ell\pm i(\mathfrak p\pm\mathfrak r)\right) }{ \Gamma(4\ell)}\, e^{-\frac{\beta}{2C}\mathfrak r^2} e^{\frac{\sigma}{C} \left(\mathfrak r^2 -\mathfrak p^2\right)}~. \label{eq.FZZTFZZT}
\end{align}
Above, we have introduced rescaled $\hat s,\hat s' = s\, b, s'\, b$ parameters for both the FZZT branes. Lastly, the saddle point evaluation of \eqref{eq.FZZTFZZT} in the $C\to\infty$ limit gives,
\begin{equation}
	\langle \FZZT|\mathbb V_\ell(x,\bar x)|\FZZT\rangle \sim \left( \frac1{2\sinh\left(2\pi\hat s+ \frac{2\pi}\beta \left(\hat s'-\hat s-i\right) \sigma\right)} \right)^{4\ell}~.
\end{equation}

\section{Discussion}

\begin{figure}[tbp]
\begin{center}
\includegraphics[width=0.8\textwidth]{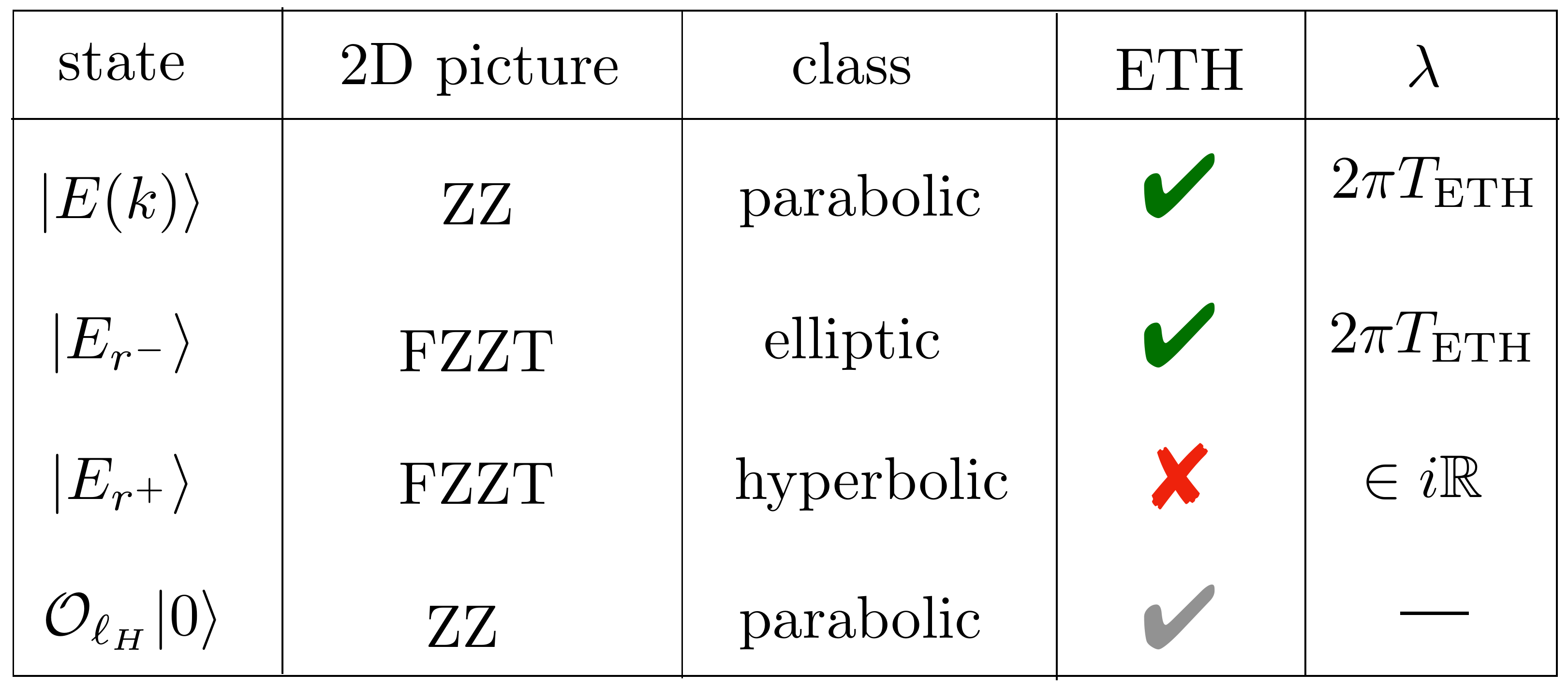}
\caption{Panoramic summary of the pure states studied in this paper and their thermal properties. We see that in all cases where the system behaves thermally it satisfies maximal extended ETH (we did not compute the Lyapunov exponent in the last line of this table). The states $|E_{r^\pm}\rangle$ denote the coherent FZZT states for $r \,\, {}^>_< \,\, \sqrt{2}$. The grey tick mark indicates that the states in question show only a weak form of ETH.}
\label{fig.SummaryTable}
\end{center}
\end{figure}

We have undertaken a detailed study of the behavior of pure states of the Schwarzian theory as defined by the path integral   \eqref{eq.Schwarz}. A table showing the different classes of states considered and their thermal properties is given in Figure \ref{fig.SummaryTable}. Let us briefly review what has been established. From a many-body perspective this paper should be understood in conjunction with the numerical results of \cite{Sonner:2017hxc} and the analytical results of \cite{Nayak:2019khe} to give a complete picture of the thermal properties of pure states of the SYK model. These studies taken together  explicitly demonstrate the applicability of ETH both for the tower of massive states by studying the asymptotics of OPE coefficients in \cite{Nayak:2019khe}, as well as for the Schwarzian sector in the present work, by employing a combination of methods, all of which are ultimately linked to the Virasoro symmetry which shows up via the orbits of Diff(S$^1$). This is yet another demonstration of the usefulness of the SYK and SYK-like models as controllable theoretical laboratories of strongly coupled quantum chaotic systems.  At the same time, these models serve as simple examples of holographic duality and the applicability of ETH has important implications for the physics of the dual black holes, as discussed in section \ref{sec.SymmaryOfResults}.

It is natural to ask whether a more detailed investigation could unearth a story similar to the case of large$-c$ 2D CFT, where sub-leading corrections necessitate a more refined analysis of thermalization to the generalized Gibbs ensemble of KdV charges (see, for example \cite{Basu:2017kzo,Dymarsky:2018lhf,Datta:2019jeo,Maloney:2018yrz,Maloney:2018hdg,Dymarsky:2019etq}). To this end, we note that we have the exact results of many relevant quantities at our disposal (see section \ref{sec.ExactResults}), so one could in principle undertake a systematic expansion in large$-C$ to address this question. A second important angle is the maximal extended ETH conjecture made in \cite{Sonner:2017hxc} and in this paper: the IR limit of the SYK model as well as other similar many-body theories (such as the melonic tensor models) scramble with the maximally allowed exponent $2\pi/\beta_{\rm eff}$ predicted from the ETH as applied to simple operators. It would be interesting to further explore whether there are universal critical phenomena associated to the ergodic / non-ergodic transition established here, and linked to the critical parabolic orbit of Virasoro. A possible connection between ETH and the Lyapunov behavior was first suggested in \cite{Sonner:2017hxc} and later formalized in \cite{Foini:2018sdb}, who point out the general structure of non-Gaussianities this implies for the statistics of off-diagonal matrix elements in the energy basis. It would be of interest to compute these non-Gaussianities explicitly in the present model, but for this we would need to extend our results to off-diagonal matrix elements.

Such an expansion of the exact results would naturally be of interest also in discussion of how the correlation function deviates from semi-classical bulk EFT expectations, as discussed in various recent analyses, such as for example \cite{Iizuka:2008hg,Cotler:2016fpe,Anous:2016kss,Fitzpatrick:2016ive}.

It is interesting to note that an ergodic to non-ergodic phase transition was also found in \cite{Kourkoulou:2017zaj, Dhar:2018pii} within pure states of the SYK model. However, the connection with the pure states of the above paper is  presently not clear and begs a more detailed analysis.

An interesting aspect with potentially important ramifications for holography is the close connection between ETH and approximate quantum error correction pointed out in  \cite{Brandao:2017irx}. Combining their results with what has been established in this work,  implies as a corollary that heavy eigenstates of SYK-like models (more precisely their IR Schwarzian limit) form an approximate quantum error correcting code (AQECC). It would be very interesting to investigate what more can be said about the properties of the AQECC hosted by the eigenstates of SYK-like models in the light of  the OTOC version of ETH we found. Recent investigations linking ETH to AQECC in chaotic theories, including holographic ones, have appeared in  \cite{Bao:2019bjp}, while \cite{Balasubramanian:2019wgd} link  complexity of time evolution to ETH-type behavior.

An alternative approach to relate spectral statistics and Lyapunov behavior is developed in  \cite{Hanada:2017xrv, Gharibyan:2018fax, Gharibyan:2019sag} who introduce new measures of quantum chaos defined in terms of the random nature of matrices of correlation functions of all the operators in a system. Given the exact results developed in this work, it would be interesting to attempt compute these measures in the Schwarzian theory.

We would like to end by reiterating an important remark about our results demonstrating ETH. For a complete proof of eigenstate thermalization in a system it is indispensable to demonstrate the exponential suppression of the off-diagonal matrix elements with respect to the diagonal terms. This was addressed numerically in \cite{Sonner:2017hxc}, as well as analytically in the conformal limit in \cite{Nayak:2019khe}. Our present work focused on the study of  the diagonal terms only, so it will be important to generalize our techniques to allow the determination of the behavior of the off-diagonal terms.

\section*{Acknowledgments}
We would like to thank Tarek Anous, Tolya Dymarsky, Daniel Jafferis, Kristan Jensen and Thomas Mertens for very helpful conversations on the subject of this paper. This work has been supported by the Fonds National Suisse de la Recherche Scientifique (Schweizerischer Nationalfonds zur F\"orderung der wissenschaftlichen Forschung) through Project Grants 200021 162796 and 200020 182513 as well as the NCCR 51NF40-141869 “The Mathematics of Physics” (SwissMAP). PN acknowledges support from the College of Arts and Sciences of the University of Kentucky.
%%%%%%%%%%%%%%%%%%%%%%%%%%%%%%%%%%%%%%%%%%%%%%
%%%%%%%%%%%%%%%%%%%%%%%%%%%%%%%%%%%%%%%%%%%%%%
%%%%%%%%%%%%%%%%%%%%%%%%%%%%%%%%%%%%%%%%%%%%%%
%%%%%%%%%%%%%  	NEW SECTION BEGINS 		%%%%%%%%%%%%%%
%%%%%%%%%%%%%%%%%%%%%%%%%%%%%%%%%%%%%%%%%%%%%%
%%%%%%%%%%%%%%%%%%%%%%%%%%%%%%%%%%%%%%%%%%%%%%
%%%%%%%%%%%%%%%%%%%%%%%%%%%%%%%%%%%%%%%%%%%%%%

\bibliographystyle{utphys}
\bibliography{refs3}{}

\end{document}